\documentclass[authoryear,preprint,review,12pt]{elsarticle}

\usepackage{amssymb}

\usepackage{amsmath,graphicx, bm}
\usepackage[dvips]{epsfig}
\usepackage{subfigure}
\usepackage{natbib}
\usepackage{boldline, multirow}
\bibpunct{(}{)}{;}{a}{,}{,}
\usepackage{color}
\usepackage{lineno,hyperref}
\usepackage{tabularx}
\usepackage[flushleft]{threeparttable}
\usepackage{booktabs}

\usepackage{algorithm}
\usepackage{algpseudocode}

\usepackage{hyperref}
\hypersetup{
    bookmarks=true,         
    unicode=false,          
    pdftoolbar=true,        
    pdfmenubar=true,        
    pdffitwindow=true,      
    pdftitle={},    
    pdfauthor={},     
    pdfsubject={},   
    pdfnewwindow=true,      
    pdfkeywords={}, 
    colorlinks=true,       
    linkcolor=blue,          
    citecolor=blue,        
    filecolor=blue,      
    urlcolor=blue           
}

\journal{xxx}

\begin{document}

\begin{frontmatter}



\title{Model predictive control and moving horizon estimation for adaptive optimal bolus feeding in high-throughput cultivation of \textit{E. coli}}


\author[First, Second]{Jong Woo Kim \corref{foot}}
\author[First]{Niels Krausch}
\author[First]{Judit Aizpuru}
\author[Third]{Tilman Barz}
\author[Fourth]{Sergio Lucia}
\author[First]{Peter Neubauer}
\author[First,Fifth]{Mariano Nicolas Cruz Bournazou}

\address[First]{Technische Universit{\"a}t Berlin, Chair of Bioprocess Engineering, Strasse des 17. Juni 135, 10623 Berlin, Germany}
\address[Second]{Department of Energy and Chemical Engineering, Incheon National University, 22012 Incheon, Republic of Korea.}
\address[Third]{AIT Austrian Institute of Technology GmbH, Center for Energy, Giefinggasse 4, 1210 Vienna, Austria}
\address[Fourth]{Technische Universit{\"a}t Dortmund, Department of Biochemical and Chemical Engineering, Emil-Figge-Strasse 70, 44227 Dortmund, Germany}
\address[Fifth]{DataHow AG, Z{\"u}richstrasse 137, 8600 D{\"u}bendorf, Switzerland}

\cortext[foot]{Corresponding author, E-mail: jong.w.kim@tu-berlin.de}

\begin{abstract}
We discuss the application of a nonlinear model predictive control (MPC) and a moving horizon estimation (MHE) to achieve an optimal operation of \textit{E. coli} fed-batch cultivations with intermittent bolus feeding. 24 parallel experiments were considered in a high-throughput microbioreactor platform at a 10 mL scale. The robotic island in question can run up to 48 fed-batch processes in parallel with automated liquid handling and online and at-line analytics. The implementation of the model-based monitoring and control framework reveals that there are mainly three challenges that need to be addressed; First, the inputs are given in an instantaneous pulsed form by bolus injections, second, online and at-line measurement frequencies are severely imbalanced, and third, optimization for the distinctive multiple reactors can be either parallelized or integrated. We address these challenges by incorporating the concept of impulsive control systems, formulating multi-rate MHE with identifiability analysis, and suggesting criteria for deciding the reactor configuration. In this study, we present the key elements and background theory of the implementation with \textit{in silico} simulations for bacterial fed-batch cultivation. 
\end{abstract}




\end{frontmatter}




\section{Introduction} \label{sec:Intro}
Mathematical descriptions of the dynamics of the microbial systems used in biomanufacturing are key to exploiting bioprocess automation \citep{leavell2020high}. From the early screening phases to the industrial-scale reactor, a consistent model-based approach helps to accelerate product development \citep{neubauer2013consistent}. The mathematical description of the mechanistic model allows for performing the optimization, which leads to the cost-effective, consistent, and highly confident bioprocess design. In this context, there has been significant progress related to the model-based bioprocess automation, such as structuring a mechanistic model that describes key metabolic pathways, parameter estimation and design of experiment, optimal control, and modularized computational frameworks \citep{gomes2015integrating, narayanan2020bioprocessing, hemmerich2021pyfoomb, herwig2021digital}.

One of the distinguishing characteristics of biological systems is that the growth of the microorganism exhibits a high level of uncertainty. Moreover, the most used cultivation strategy is fed-batch, in which no steady-state exists and the process variables vary over a wide range. Hence, a large amount of data is needed to obtain a mathematical model which is capable of describing the non-steady-state dynamics of highly nonlinear and complex systems such as cells \citep{cruz2017online}. The recent development of high-throughput (HT) technology allows for performing a large number of laborious and time-consuming experiments by automatizing, parallelizing, and miniaturizing the experimental facilities \citep{puskeiler2005development, bunzel2018speeding, hemmerich2018microbioreactor}. Liquid handling stations support the parallel cultivation of a large number of mini-bioreactors, based on the automated sampling, on-line and at-line analytics operation, real-time control, and data acquisition \citep{kusterer2008fully, tai2015efficient, haby2019integrated, hans2020monitoring}. HT technology plays an important role, especially in the early stage of bioprocess development. Very large amounts of data from testing the cell clones under different conditions such as media, pH, temperature, induction, and feeding strategies become available \citep{hans2020automated} allowing us to understand the process better and to find the best-operating conditions in very large search spaces. This has naturally led to the enhancement of the conditional screening and strain phenotyping \citep{schmideder2016parallel, sawatzki2018accelerated, janzen2019implementation, fink2021high}.

To exploit the full potential of the HT bioprocess development (HTBD), it is essential to couple the model-based methods with robotic facilities. The reliability of computed decisions depends on the accuracy of model predictions, e.g., each cell type and organism needs tailored feeding strategies for optimal growth or product formation. Because of limited reproducibility at the small-scale reactors at the $\mu$L and mL scale, these strategies need to be adaptively controlled. In addition, monitoring of cultivations is often challenging because of the limited accuracy of online sensors, e.g., small delays and shifts in oxygen signals, or large delays in the processing of liquid samples. Finally, the satisfaction of experimental constraints is critical, such as minimum levels of sugar and/or dissolved oxygen concentrations to minimize specific stresses, i.e., mainly glucose excess, limitation and starvation, and oxygen exhaustion \citep{delvigne2009bioreactor}. 

In previous works, intermittent-feeding strategies have been derived from established exponential feeding strategies considering the (previously or online identified) maximum specific growth rate of the organism \citep{sawatzki2018accelerated, anane2019modelling, hans2020automated}. Because of the limited information and feedback, it has been observed that this approach can lack robustness in real applications. There are only a few studies that incorporate model-based methods for HT experiments. However, the studies consider the optimal experimental design for the model fitting instead of optimal operation for maximizing biomass \citep{cruz2017online, barz2018adaptive, kim2021oed}.

Model predictive control (MPC) and moving horizon estimation (MHE) are popular methods in engineering \citep{rawlings2017model} and there has been a large number of applications for fed-batch bioreactor cultivations. Given the state trajectory from the pre-defined operating strategy, MPC has been applied for the optimal tracking control of the bioprocesses described with basic Monod equations \citep{ramaswamy2005control, tebbani2008open}, and further extended to economic objectives such as to maximize the product \citep{ashoori2009optimal, raftery2017economic}. In the presence of measurement and model uncertainties, optimal state or parameter estimators such as Kalman filter \citep{markana2018multi} or MHE \citep{abdollahi2012lipid, del2016model} are combined with MPC in order to adapt to the data. Various optimization methods have been studied such as maximization principle \citep{pvcolka2016algorithms, luna2017iterative} and evolutionary strategies \citep{freitas2017optimization}. As an alternative to MPC, the reinforcement learning (RL) method has been studied to reduce the online computation time by obtaining the closed-loop policy and for optimal operation even without the mechanistic model \citep{martinez2013dynamic, petsagkourakis2020reinforcement, kim2021model}. 

Although several studies have proven that the model-based approach comprising a controller and estimator is promising for automated bioprocess operation \citep{lucia2017adaptive}, it has, to our knowledge, never been implemented in parallel mini-bioreactors for HTBD. The extension to this application is not trivial and has important challenges that can lead to poorly operated experiments. We discuss the most relevant challenges and how they have been tackled in the presented framework. First, inputs are given in a pulse form by bolus injections. In highly parallelized milliliter scale mini-bioreactors for example, individual feeding with high accuracy is very challenging and costly \citep{faust2014feeding}. As an alternative, pulse-based feeding has been widely utilized. In contrast to continuous feeding, intermittent feeding fails to achieve high cell density cultivation, because it induces large heterogeneities in the operating conditions such as oscillating pH, oxygen, temperature, glucose concentrations, and toxic compounds \citep{neubauer2013consistent}. This can however when properly designed and operated using scale-down techniques, be used to mimic the industrial scale heterogeneity \citep{neubauer2016scale}. According to \citet{anane2019modelling}, the combination of a pulse-based scale-down approach is successful to test strain robustness and physiological constraints at the early stages of bioprocess development. The pulse-based (bolus) feeding is different compared to the standard process control problems where systems have continuous input and dynamics. This requires a different approach, namely, impulsive control systems  \citep{yang2001impulsive}. 

Second, the measurements are multi-rate, the sampling times between the parallel bioreactors are not aligned, and the sample delays are on the order of tens of minutes such that a multi-rate formulation of the MHE is necessary. The system is limited to extracellular measurements, which makes it very difficult to generate sufficient information to properly identify mechanistic models that aim to describe a highly nonlinear and large biochemical network. For this reason, efficient methods to assure a well-posed state and parameter estimation problem are needed. 

Third, given the configuration of multiple cultivation conditions and bioreactors, the proper design of the experimental campaign and distribution of the operating strategies are not trivial. The information in parallel systems (e.g. replicates) needs to be efficiently exploited to improve the probability of finding the optimal process condition.

Our contribution focuses on the computational model-based framework, namely, MPC and MHE for the optimization of the cultivation conditions for maximal growth of \textit{E. coli} in parallel mini-bioreactors. Three bottlenecks encountered in the HT experiment, bolus injection, multi-rate measurements, and multiple bioreactors, are addressed: First, we illustrate how to design the objective and constraint function for the impulsive control systems. Unlike the zero-order hold discretization which has been used for impulsive systems, a full-discretization method is implemented to capture the fast dynamics happening within the pulse-feed interval. Second, we propose an arrival cost design and identifiable parameter selection method for the multi-rate and partially ill-conditioned MHE. Third, for the purpose of determining the reactor configuration for the MHE optimization, three criteria, size of the identifiable parameter subset, root mean squared error of the MHE, and computational time are suggested. The developed framework is validated through \textit{in silico} studies on the fed-batch cultivation of \textit{E. coli}. As a proof-of-concept, we design 24 parallel cultivation experiment, which is divided into 8 different conditions with 3 replicates. We focus on whether the MPC and MHE are capable of providing a consistent/feasible pulse-feeding strategy regardless of the operating condition. The experimental validation is conducted in the companion paper \citep{krausch2022htbd}.

The remainder of this paper is organized as follows: Section 2 describes the \textit{in silico} experiments of mini-bioreactors and the challenges of the HT experiment. Section 3 introduces model-based methods, MHE and MPC. In Section 4, the simulation results are discussed. Finally, a discussion and some concluding remarks are provided in Section 5.

\section{\textit{In silico} experiments on a parallel mini-bioreactor platform}
\subsection{Experimental setup, sensors, and automated liquid handling}
\label{sec:MBR}
The robotic facility of the HTBD platform is able to conduct 24 parallel cultivations. The liquid handling station assists in sampling, measuring, glucose feeding, medium balancing, and pH control (acid and base). The 24 mini-bioreactors are placed in three columns and eight rows, and we put a numeric order to the columns (i.e., 1, 2, and 3), and an alphabetical order to the rows (i.e., A, B, ..., H). Based on this configuration, reactors that have the same columns are replicates and the HT experiment is conducted with eight different cultivation conditions. 

The purpose of the study is to see how the proposed MPC and MHE scheme adaptively computes the feeding strategy for various experimental conditions. To see this, this study tests the effect of three factors on the feeding strategy and cell growth, changing two conditions for each factor. 
Three factors are chosen as (1) type of strain (strain I and II), each has unique characteristics which are modeled by different parameter sets I and II of the growth model presented in Section~\ref{sec:MKGmodel}, (2) initial biomass/glucose concentrations, and (3) the lowest DOT bound for bioreactors. The experimental design is described in Table~\ref{tb:MBR8condns}.

\begin{table}
	\caption{Experimental design of the HT experiment. Each reactor row consists of three replicates.}
	\begin{center}
		\begin{tabular}{c c c c c} \toprule
	\multirow{2}*{Reactor row} & \multirow{2}*{Strain}	 & \multicolumn{2}{c}{Initial condition ($g/L$)} & \multirow{2}*{DOT lower bound ($\%$)}  \\
	\cmidrule(lr){3-4}
   &  & Biomass & Glucose & \\ \hline
A & I & 3.55 & 0.25 & 20  \\
B & I & 3.55 & 0.25 & 10  \\
C & I & 2.86 & 0.16 & 20   \\
D & I & 2.86 & 0.16 & 10   \\
E & II & 3.55 & 0.25 & 20   \\
F & II & 3.55 & 0.25 & 10  \\
G & II & 2.86 & 0.16 & 20  \\
H & II & 2.86 & 0.16 & 10  \\
\hline 
		\end{tabular} 
	\end{center}
	\label{tb:MBR8condns}
\end{table}

\begin{figure}
\centering
\includegraphics[width=\linewidth]{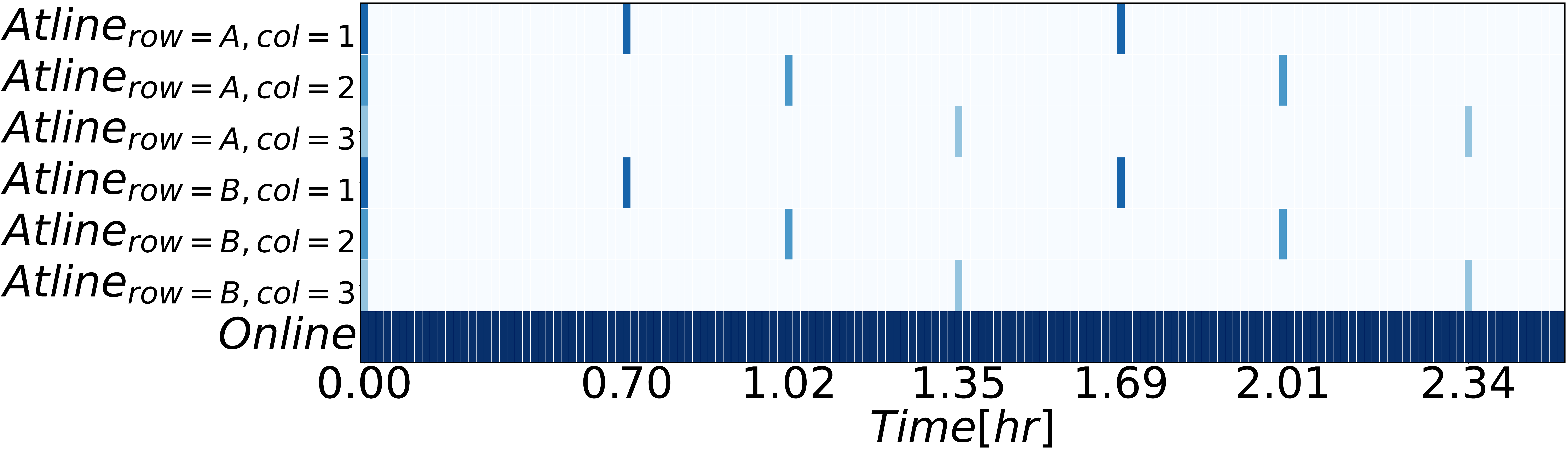} 
\caption{Atline and online measurements from the three replicates (i.e., columns 1, 2, and 3) of the reactors in rows A and B. The graph is colored only when measurements are available at the time.}
\label{fig:triplicate_meas}
\end{figure}

The following measurements and control actions are considered.
\begin{itemize}
    \item Online measurements: DOT and pH are recorded every 30 seconds online. 
    \item Atline measurements: Biomass, substrate, acetate, and product concentrations are measured from the sample every 60 minutes from a single replicate. The sample is obtained from a single replicate alternately with 20 minutes time differences; The second and third replicates are measured after 20 and 40 minutes after the first replicate, respectively. The initial concentrations are measured for all replicates.
    \item Bolus feeding: Glucose solution 200 $g/L$ concentration is added every 10 min to each reactor. The amount of glucose pulses is the decision variable of the MPC. The pH of each mini-bioreactor is controlled by the intermittent addition of the acid and base solutions. The pH is assumed constant. The culture medium is added to compensate for the volume reduction due to the sampling and evaporation. The medium addition takes place in between the volume additions from the pH controller and glucose feeding.
\end{itemize}
Figure~\ref{fig:triplicate_meas} illustrates how at-line and online measurements are obtained from the three replicates (i.e., columns 1, 2, and 3) of the reactors in rows A and B. For each replicate, at-line analytics are measured with 20 minutes offset between each other. DOT is measured every 30 seconds for all reactors. The at-line measurements from the same column share the equivalent sample time.

According to \citet{haby2019integrated}, the liquid handling is scheduled to perform the sterilization of needles, volume balance, pH control, Feeding, sample collection, at-line sampling, and off-line sampling at each time. The samples are injected into the microplate, together with the baseline solutions. To analyze the at-line samples triplicates, we do not sample from the triplicates together, instead sample from a single reactor at each time, and sample alternately. Because there exists a time duration for the at-line analytics to be performed, the frequency of at-line measurements is significantly lower than that of on-line measurements. Note that due to the sensor delay, data processing times, limited transfer times and dispense speed of the liquid handling, the measurements are not strictly regular in reality. Detailed descriptions about the robotic facility and data handling processes are provided in \citep{haby2019integrated}. The reader is also referred to \citep{krausch2022htbd} for experimental validation following the in silico results including information on the strains and sampling/analytics.

A typical bioprocess can be divided into 3 phases: the batch phase, a subsequent feeding phase to achieve high cell densities, and finally an induction phase in which a recombinant protein is produced by the cells after the addition of an inducing agent. The end of the batch phase is characterized by the complete depletion of the substrate and an associated sudden increase in DOT. In the feeding phase, the substrate is either fed in an exponential manner for a fixed period of time or until a predefined cell concentration has been reached. Subsequently, production of the recombinant protein is started by the addition of a promoter-specific inducer and a constant feed is applied, as the cells are exposed to an increased stress level during this phase.

\subsection{Macro-kinetic \textit{E. coli} growth model} \label{sec:MKGmodel}
The macro-kinetic growth model consists of a set of ordinary differential equations (ODEs) with six state variables: biomass $X$, substrate $S$, acetate $A$, dissolved oxygen tension measurement $DOT_m$, product $P$, and reactor medium volume $V$. The governing equations for $X$, $S$, $A$, $P$, and $V$ are expressed as
\begin{align} \label{eq:MKG_concentrations}
    \dfrac{dX}{dt} &= -\mu X - \dfrac{X}{V} \cdot \dfrac{dV}{dt} \\
    \dfrac{dS}{dt} &= -q_S X - \dfrac{S}{V} \cdot \dfrac{dV}{dt} \\ 
    \dfrac{dA}{dt} &= q_A X - \dfrac{A}{V} \cdot \dfrac{dV}{dt} \\
    \dfrac{dP}{dt} &= q_P X - \dfrac{P}{V} \cdot \dfrac{dV}{dt} \\
    \dfrac{dV}{dt} &= -F_{evap}
\end{align}
where $\mu$, $q_S$, $q_A$, and $q_P$ are the specific growth rate ($g/(g\cdot h)$), specific substrate uptake rate ($g/(g\cdot h)$), specific acetate production rate ($g/(g\cdot h)$), and specific product formation rate ($g/(g\cdot h)$), respectively; $F_{evap}$ ($L/h$) is the evaporation rate. The detailed model description is provided in Appendix~\ref{Apx:MKGmodel}. Dissolved oxygen tension (DOT) is modeled by the differential equation as
\begin{equation}
\dfrac{dDOT}{dt} = k_{la}(DOT^* - DOT) - q_OXH \label{eq:MKG_DOTdiff}
\end{equation}
where $k_{la}$ ($h^{-1}$) denotes the volumetric oxygen transfer coefficient; $DOT^*$ ($\%$) denotes the saturation concentration of DOT; $q_O$ ($g/(g\cdot h)$) denotes the specific oxygen uptake rate (see Eq.~(\ref{eq:MKG_qO})); $H$ ($mol/(m^3 \cdot Pa)$) denotes the Henry constant. Considering that DOT has fast dynamics \citep{duan2020model}, we use the reduced form expressed in the algebraic equation as
\begin{equation} \label{eq:MKG_DOTalg}
DOT = DOT^* - \dfrac{q_OXH}{k_{la}}
\end{equation}
The delayed response of the DOT measurement $DOT_m$ is described by the first-order dynamics as
\begin{equation} \label{eq:MKG_DOTm}
\dfrac{dDOT_m}{dt} = k_{p}(DOT - DOT_m)
\end{equation}
where $k_p$ ($h^{-1}$) represents the time constant.

\begin{table}
	\caption{Parameters of macro-kinetic growth model}
	\begin{center}
		\begin{tabular}
		{l l c c c} \toprule
 \multirow{2}*{Param.} & \multirow{2}*{Description} & \multicolumn{2}{c}{Value} & \multirow{2}*{Unit} \\ \cmidrule(lr){3-4} 
 & & Strain I & Strain II &\\ \hline
$q_{S, max}$ & Max. specific substrate uptake rate & 1.65 & 1.52 & $g/(g\cdot h)$ \\
$q_m$ & Specific maintenance coefficient & 0.044 & 0.044 & $g/(g\cdot h)$ \\
$q_{Ap, max}$ & Max. specific intracell. uptake rate for acetate flux & 0.62 & 0.30 & $g/(g\cdot h)$ \\
$q_{Ac, max}$ & Max. specific acetate consumption rate & 1.46 & 0.60 & $g/(g\cdot h)$ \\
$Y_{XS,em}$ & Yield (biomass/substrate) excluding maintenance & 0.62 & 0.54 & $g/g$ \\
$Y_{AS,of}$ & Yield (acetate/substrate) in overflow metabolism & 0.667 & 0.667 & $g/g$ \\
$Y_{XA}$ & Yield (biomass/acetate) & 0.20 & 0.20 & $g/g$ \\
$Y_{OS}$ & Yield (oxygen/substrate) & 1.23 & 1.23 & $g/g$ \\
$Y_{OA}$ & Yield (oxygen/acetate) & 0.52 & 0.52 & $g/g$ \\
$Y_{PS}$ & Yield (product/substrate) & 0.10 & 0.10 & $g/g$ \\
$K_{S}$ & Affinity for substrate consumption & 0.069 & 0.010 & $g/L$ \\
$K_{qS}$ & Affinity to intracell. substrate flux & 5.8 & 1.19 & $g/(g\cdot h)$ \\
$K_{i,SA}$ & Inhibition of substrate uptake by acetate & 1.86 & 0.50 & $g/L$ \\
$K_{A}$ & Affinity for acetate consumption & 0.29 & 0.05 & $g/L$ \\
$K_{i,AS}$ & Inhibition of acetate uptake by substrate & 1.21 & 0.50 & $g/L$ \\
$d_{S,ox, P}$ & Distribution ratio of oxidative flow for product & 0.60 & 0.60 & $\%$ \\\hline
		\end{tabular} 
	\end{center}
	\label{tb:MKG_global_params}
\end{table} 

\begin{table}
	\caption{Reactor-dependent parameters of macro-kinetic growth model}
	\begin{center}
		\begin{tabular}
		{l l l l} \hline
 Param. & Description & Value range & Unit \\ \hline
$k_{la}$ & Volumetric oxygen transfer coefficient & [450, 800] &  $h^{-1}$ \\
$k_p$ & Response time of the oxygen sensor & [50, 80] & $h^{-1}$ \\ \hline
		\end{tabular} 
	\end{center}
	\label{tb:MKG_local_params}
\end{table} 

\begin{table}
	\caption{Constants of macro-kinetic growth model}
	\begin{center}
		\begin{tabular}{l l l l} \hline
  Constant & Description  & Value & Unit \\ \hline
$H$ & Henry constant & $1.4\times 10^4$ & $mol/(m^3 \cdot Pa)$ \\
$DOT^*$ & DOT at saturation & 100 & $\%$ \\
$F_{evap}$ & Evaporation rate & $5.0\times 10^{-5}$& $L/h$  \\
$S_f$ & Substrate concentration in the feed & 200 & $g/L$  \\
\hline 
		\end{tabular} 
	\end{center}
	\label{tb:MKGconsts}
\end{table}

Table~\ref{tb:MKG_global_params} provides the parameter set I and II of the two strains which describe significantly different cell metabolisms. The macro-kinetic growth model and its parameter set I are adjusted from  \citet{anane2017modelling}. Parameter set II is an artificial value that shows a different behavior from that of parameter set I. In addition, Table~\ref{tb:MKG_local_params} provides reactor-related parameters, $k_{la}$ and $k_p$, which characterize the mixing and oxygen transfer in each reactor. We assume that each reactor has random values within the range given in Table~\ref{tb:MKG_local_params}. Constant values and their units are listed in Table~\ref{tb:MKGconsts}. 

\subsection{Characteristics of the high-throughput experiment from an optimal control perspective}
\subsubsection{Impulsive control systems} \label{sec:ImpulsiveControlSys}
The injection in the intermittent feeding phase results in sudden concentration and volume changes. These changes are modeled as instantaneous changes. Thus, the state trajectory shows discontinuities (jumps) in response to the pulse feed. Accordingly, the studied system to which intermittent/pulse feeding is applied is characterized as an impulsive control system \citep{yang2001impulsive}. Impulsive systems appear in various industrial applications such as spacecraft, communication systems, dosage supply in pharmacokinetics, and fed-batch bioprocesses \citep{yang2019recent, villa2020adaptive, shen2008bilevel}. Unlike continuous control cases, there exist manipulated variables that are given only in discrete times, hence imposing additional discrete time sets and difference equations. Moreover, due to the quasi-instantaneous change of state variables originating from the impulsive control, new stability properties and algorithms are required \citep{yang2019recent}. In \citep{sopasakis2014model} the stability properties of the impulsive linear system are studied. An MPC for the impulsive system which uses a modified cost function is developed \citep{rivadeneira2017control, villa2020adaptive}.

\begin{figure}
\centering
\includegraphics[width=\linewidth]{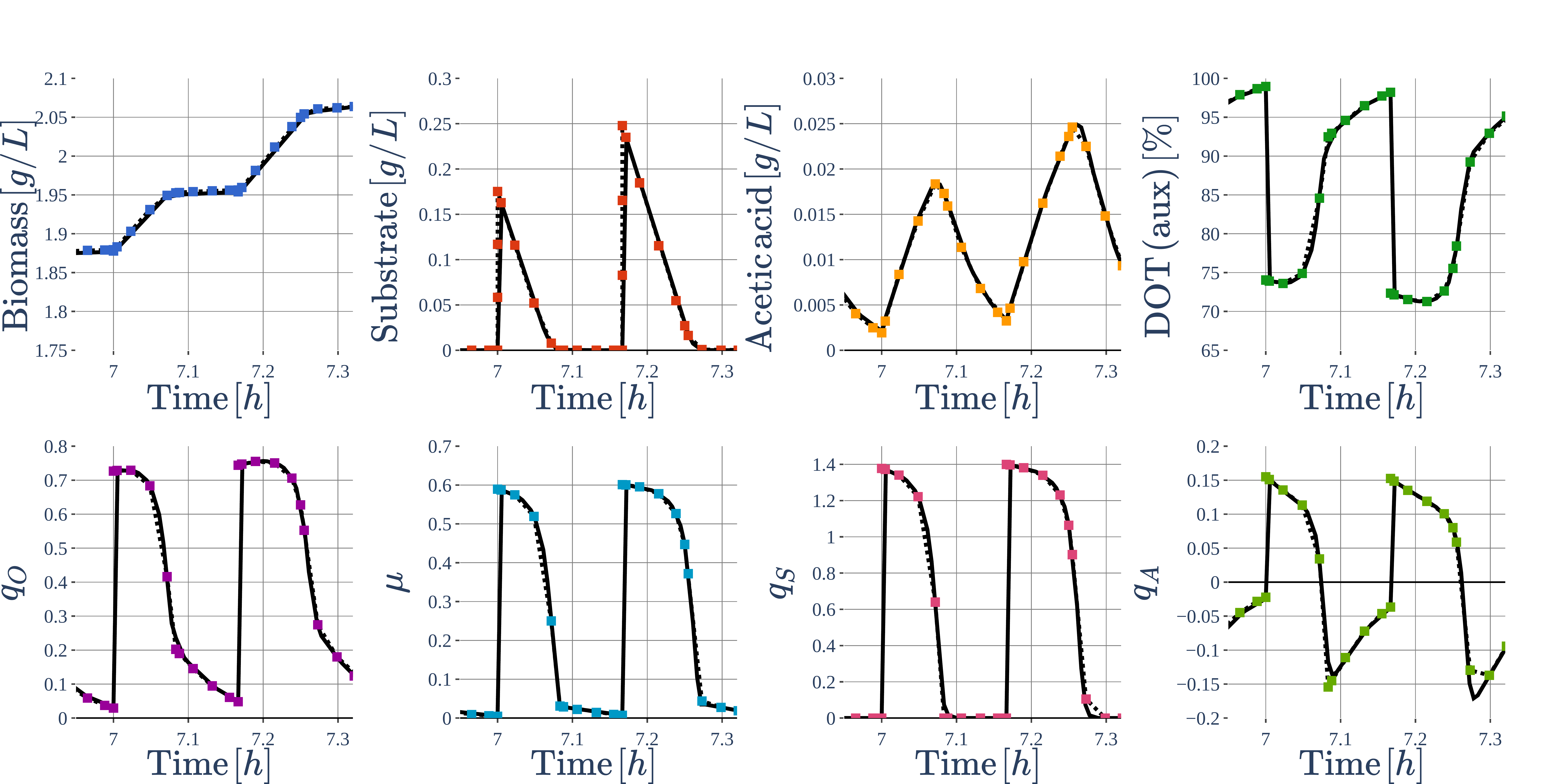} 
\caption{The states (i.e., $X$, $S$, $A$, and $DOT$) and specific rates (i.e., $q_O$, $\mu$, $q_S$, and $q_A$) during the two pulse-feeding periods. The solid line represents the simulation and the dotted line with marker represents the collocation points obtained from solving the optimization problem in Eq.~(\ref{eq:CultOptProb}).}
\label{fig:Colloc_high_resolution}
\end{figure}

Generally speaking, the dynamics of the impulsive control systems can be transformed into algebraic equations by discretizing the differential equations. Previous studies for the impulsive MPC simply use the zero-order-hold discretization \citep{sopasakis2014model, rivadeneira2017control, yang2019recent, villa2020adaptive}. This causes an \textit{aliasing} effect to the MPC, in which the discretized impulsive state-space model can only consider the state values at the discrete-time set where impulsive inputs are given. Nonetheless, this is inappropriate for our case, since the state dynamics in between the pulse-feed interval are important in terms of estimation and constraint violation. The pulse-feeding induces the instantaneous perturbation to the states and the growth parameters and the values are quickly recovered back to the pseudo-steady-states where zero feed is implemented. This phenomenon is illustrated in Fig.~\ref{fig:Colloc_high_resolution}. Furthermore, in order to capture and distinguish the fast and slow dynamics, we use the full-discretization method for formulating the optimization problem of the MPC. This means that pulse feeds are considered by `explicit (or time) events' which trigger updates of state variables \citep{jouned2022event}. It is noted that state changes due to sampling (volume) and additions for pH control (volume and concentration) can be considered in the same way. The discretization scheme is aligned with the location of these events. The updates are obtained from the mass balance equations considering the reactor states before ($t$) and after ($t^+$) the feed injection. They are computed as
\begin{subequations} \label{eq:MKG_MB}
\begin{align}
X(t^+) &= X(t) - \dfrac{X(t)}{V(t^+)}\left(\Delta v(t) + V_a(t) + V_s(t) \right)  \\
S(t^+) &= S(t) - \dfrac{S(t) - S_f}{V(t^+)} \left(\Delta v(t) + V_a(t) + V_s(t) \right) \\
A(t^+) &= A(t) - \dfrac{A(t)}{V(t^+)}\left(\Delta v(t) + V_a(t) + V_s(t) \right) \\
DOT_m(t^+) &= DOT_m(t) \\
P(t^+) &= P(t) - \dfrac{P(t)}{V(t^+)}\left(\Delta v(t) + V_a(t) + V_s(t) \right) \\
V(t^+) &= V(t) + \Delta v(t) + V_a(t) + V_s(t) 
\end{align}
\end{subequations}
where $S_f$ ($g/L$) is the substrate concentration in the feed; $V_a$ and $V_s$ ($L$) are the amount of volume addition and subtraction from the pH control, medium balancing, and sampling, respectively.

Due to the existence of another discrete-time set consisting of the pulse-feeding times, the cost function and constraint should be modified. To tackle this, we perform additional analysis on shaping the cost function and defining the constraint in Section~\ref{sec:MPCformulation}.

\subsubsection{Multi-rate measurements and ill-conditioning of the estimation problem}
The measurements in the HT experiment possess several characteristics that hamper the performance of the state and parameter estimation. First, according to Fig.~\ref{fig:triplicate_meas}, the at-line and online measurement frequencies are considerably imbalanced. Second, the sampling times are irregular, delayed, and not aligned. In other words, the low-frequent at-line and high-frequent online sample times  do not coincide. Third, only a subset of the states is measurable, for example, $DOT$ is not directly measured. The estimator for the HT experiment must address such challenges.

State and parameter estimation methods can be classified into recursive methods and optimization-based methods, namely Kalman filter (KF) and MHE, respectively \citep{alexander2020challenges}. Most of the estimators follow the assumption that online measurements are available for every time instance and desired state variable. Multi-rate estimators have been studied in the bioprocess in an attempt to make use of both fast and slow measurements. MHE has been applied to the multi-rate system due to its flexibility in placing the measurements to any time instances in the optimization problem \citep{lucia2017adaptive}. Several different structures have been proposed for the multi-rate MHE; Some approaches align the two measurements by filling in empty values for the slow measurement with a zero-order hold method \citep{kramer2005fixed}, or a model prediction \citep{liu2016moving}. In contrast, in a variable structure only existing slow measurements are considered \citep{kramer2005multirate}. The approach of the distributed MHE regards the fast measurements from the fast dynamics as algebraic equations and solves the MHE only for the slow subsystem \citep{yin2017distributed}. In our case, the imbalance is so severe, that the time alignment might deteriorate the performance because the estimation is biased toward either the noisy measurement or the uncertain model. Moreover, the fast measurement (i.e., $DOT_m$) does not present fast dynamics. Therefore, we use the variable structure method.

Regardless of the structure, an arrival cost highly affects the performance of MHE. MHE uses only a limited window of recent measurements in order to prevent the model from being overfitted to the uncertain data. Here, arrival cost is related to the uncertainty of the unconsidered past measurements (i.e., marginalized covariance) evaluated at the previously estimated values. A proper assignment of the arrival cost helps to maintain the performance of the full horizon estimation and to compensate for the inaccuracy originating from the uncertain data. In the context of the multi-rate MHE, a family of filtering methods such as extended KF \citep{rao2002constrained}, unscented KF \citep{qu2009computation}, QR decomposition \citep{diehl2006schnelle} are utilized for the arrival cost computation. Nonlinear programming sensitivity is used to approximate the smoothed covariance with the inverse of the Hessian of the objective function \citep{lopez2012moving, fiedler2020probabilistic}. A comprehensive review of the arrival cost and stage cost for the variable structure multi-rate MHE can be found in \citep{elsheikh2021comparative}.

Joint estimation of the states and parameters may provide an unsatisfactory result, especially under the scarcity of the measurement. Recall that among six state variables, five states rely on the slow at-line measurement. This gives rise to the loss of identifiability of some parameters of the macro-kinetic growth model, which eventually leads to the failure of the MPC \citep{barz2016real}. Parameter subset selection is a method to draw a distinction between the non-identifiable parameters, which are excluded to recover from the ill-conditioned estimation problem \citep{lopez2015nonlinear}. This method can be applied to the multi-rate estimation problem by modifying the computation of the dynamic parametric sensitivity \citep{barz2018adaptive, anane2019output, bae2021multirate}. In this study, we further extend the $\epsilon$-rank subset selection method proposed in \citep{lopez2015nonlinear} to the multi-rate MHE for the impulsive control systems. We also provide the procedure to determine the corresponding arrival cost.

\subsubsection{Estimation and control for the parallel reactor systems}
Given the configuration of the 24 mini-bioreactors, the experimental design has a major influence on the performance of both, the MHE and MPC. Running parallel experiments is advantageous in alleviating the ill-conditioned estimation problem because, if properly scheduled, the samplings of several replicates can be combined to increase the information content and frequency of the data \citep{kim2021oed}. Nonetheless, the following cases might deteriorate the overall performance: First, the estimator undergoes under-parametrization where the reactors that have different kinetic parameters are integrated. Second, the online computation time exceeds the decision time interval when too many reactors are merged into a single MHE problem because the number of decision variables comprises of states to be estimated and the reactor-dependent parameters (see Table~\ref{tb:MKG_local_params}) increases proportionally. Considering the trade-off, we propose two criteria, identifiable subset size and computation time, for the decision of the configuration for the MHE. MPCs for the parallel reactors, on the other hand, are completely independent of each other. Therefore we solve 24 individual MPC problems in parallel.

\section{Model-based optimal cultivation}
\subsection{Problem description and formulation}
We describe the HT experiment and the macro-kinetic growth model in formal expressions. The states $x$, manipulated variables $u$, and measured variables $y$ comprise of 
\begin{equation}
\begin{split}
x &= \left[ X, S, A, DOT_m, P, V \right] \\
y &= \left[ X, S, A, DOT_m, P \right] \\
u &= \left[ \Delta v \right]
\end{split}
\end{equation}

The experimental procedures are characterized by the following finite sets :
\begin{equation}
\begin{split}
    \mathcal{R}&=\left\lbrace (row, col) | row \in \left\lbrace A, B, \ldots, H\right\rbrace, col \in \left\lbrace 1, 2, 3 \right\rbrace \right\rbrace \\
    \mathcal{U}&=\left\lbrace 10k \text{ (min)}| k\in \left\lbrace 0, 1, 2, \ldots \right\rbrace \right\rbrace \\
    \mathcal{M}^{r, y} &=\left\lbrace \begin{array}{l} 
        \begin{split}
            &\left\lbrace 40 + 20(i-1) + 60k \text{ (min)} | k\in \left\lbrace 0, 1, 2, \ldots \right\rbrace, r = (i, \cdot) \in \mathcal{R} \right\rbrace,\\
            &\quad y\in \left\lbrace X, S, A, P \right\rbrace
        \end{split} \\
        \left\lbrace 30k \text{ (sec)} | k\in \left\lbrace 0, 1, 2, \ldots \right\rbrace \right\rbrace, \quad  r\in \mathcal{R}, \quad y= DOT_m 
        \end{array} \right. 
\end{split}
\end{equation}
where $\mathcal{R}$ is the index set of the mini-bioreactors; $\mathcal{U}$ is the discrete pulse-feeding times; $\mathcal{M}^{r, y}$ is the measurement times of the reactor $r$ for the measured variable $y$. We assume that the feeding times are identical for all bioreactors. On the other hand, the measurement times are different for each reactor and for each measurement variable. The collection of all time elements of $\mathcal{M}^{r, y}$ is denoted as $\mathcal{M}=\bigcup_{y, r}\mathcal{M}^{r, y}$. We introduce another notation that constrains the time instance sets to an arbitrary time window $[a, b]$ as follows:
\begin{equation}
\mathcal{U}_{[a, b]} = \mathcal{U} \cap [a, b], \quad \mathcal{M}^{r, y}_{[a, b]} = \mathcal{M}^{r, y} \cap [a, b]
\end{equation}

Denote differential equations of the macro-kinetic growth model (Eqs.~(\ref{eq:MKG_concentrations})-(\ref{eq:MKG_DOTm})) as $f \in \mathbb{R}^{n_x}$, difference equations of the mass balance due to the pulse-feeding (Eqs.~(\ref{eq:MKG_MB})) as $f_d \in \mathbb{R}^{n_x}$, and output functions for reactor $r$ as $h_r \in \mathbb{R}^{n_y(t)}$. Then the impulsive control systems can be described in the compact form:
\begin{equation} \label{eq:MKGMBRCompact}
\begin{split}
\dot{x}_r(t) &= f(x_r(t), \theta_r), \quad t \in [t_0, t_f] \setminus \mathcal{U} \\
x_r(t^+) &= f_d(x_r(t), u_r(t)), \quad t\in \mathcal{U} \\
y_r(t) &= h_r(x_r(t)) \quad t \in \mathcal{M}^r, \forall y \\
x_r(t_0) &= x_{t_0, r}, \quad \forall r \in \mathcal{R}
\end{split}
\end{equation}
where the subscript $r$ indicates that variables $x$, $u$, and $y$ belong to the reactor $r$; $t_0$ and $t_f$ are the initial and final cultivation time, respectively; $t^+$ is the time after which the pulse-feed is made; $x_{0, r}$ is the initial condition of the bioreactor $r$. The parameter vector for the individual reactor $r\in \mathcal{R}$ is denoted as $\theta_r$. It contains the growth parameters $\theta_g$, the reactor-dependent parameters for the reactor $r$, $\theta_{l, r}$, and the initial state condition $x_{t_0, r}$. The collection of all parameters is denoted as $\theta$. The following shows the definition:
\begin{equation}
\begin{split}
\theta_r &= \left[\theta_g, \theta_{l, r}, x_{0, r} \right], \quad r \in \mathcal{R} \\
\theta &= \left[\theta_g, \left\lbrace \theta_{l, r} | r \in \mathcal{R} \right\rbrace, \left\lbrace x_{t_0, r} | r \in \mathcal{R} \right\rbrace \right]
\end{split}
\end{equation}
The size of the vector $\theta$ is denoted as $n_{\theta}$ is $16 + 24(2+n_x)$. The output function is given by $h_r(x_r(t)) = x_r(t)$ if $t \in \mathcal{M}^{r, y}$ and not defined elsewhere. This multi-rate nature makes the dimension of the function $h_r$ time dependent, denoted as $n_y(t)$, $t\in \mathcal{M}^{r, y}$.

Note that in the impulsive control system Eq.~(\ref{eq:MKGMBRCompact}), the manipulated variables $u_r$ are only applied to the system at $\mathcal{U}$. Accordingly, the differential equations $f$ and the difference equations $f_d$ are defined at the disjoint time sets $[t_0, t_f] \setminus \mathcal{U}$ and $\mathcal{U}$, respectively.

\subsection{Moving horizon state and parameter estimation}
\label{sec:MHE}
States and parameters are estimated using MHE. In the moving horizon approach, only a limited window of past measurements is taken into account. We denote the window $[\tau_e, \tau]$, where $\tau_e$ is the starting time of the estimation window and $\tau$ is the current time. All measurements are assumed to be independent and identically distributed and to follow a normal distribution. The residuals are weighted with the scaled diagonal variance-covariance matrix $\Sigma_{t_i}$. The diagonal elements of $\Sigma_{t_i}$ are scaled with $|\mathcal{M}^{r, y}_{[\tau_e, \tau]}|$, the number of measurements of the output $y$ of the reactor $r$ within the time window $[\tau_e, \tau]$. The multi-rate MHE is formulated for the subset of the reactors $\tilde{\mathcal{R}} \subseteq \mathcal{R}$. Then the optimization variable of the MHE is $\theta = \left[\theta_g, \left\lbrace \theta_{l, r} | r \in \tilde{\mathcal{R}} \right\rbrace, \left\lbrace x_{\tau_e, r} | r \in \tilde{\mathcal{R}} \right\rbrace \right]$, where $x_{\tau_e, r}$ denotes the state variable estimate of the reactor $r$ at time $\tau_e$. MHE is formulated as,
\begin{subequations} \label{eq:MHEobjective}
\begin{align}
\min_{\theta} &\quad \| \hat{\theta} - \theta \|_{R_0^{-1}}^2 + \sum_{r\in \tilde{\mathcal{R}}}\sum_{t_i \in \mathcal{M}_{[\tau_e, \tau]}} \left\| h_r(x_r(t_i)) - y_{r, i} \right\|_{\Sigma_{t_i}^{-1}}^2 \\
\text{s.t.}  
& \quad \dot{x}_r(t) = f(x_r(t), \theta_r), \quad t \in [\tau_e, \tau] \setminus \mathcal{U}_{[\tau_e, \tau]}  \label{eq:MHEdynamicsConstr} \\
& \quad x_r(t^+) = f_d\left(x_r(t), u_r(t)\right), \quad t\in \mathcal{U}_{[\tau_e, \tau]} \\
& \quad x_r(\tau_e) = x_{\tau_e, r}, \quad r \in \tilde{\mathcal{R}} \label{eq:MHEinitConstr}
\end{align}
\end{subequations}
where, the norm $\| x\|_P^2$ is defined as $\| x\|_P^2=x^TPx$ for an arbitrary vector with a positive definite matrix $P$; $R_0$ denotes a weighting constant for the arrival cost; $\hat{\theta}$ is an optimal estimate of $\theta$ computed at the previous decision time from solving Eq.~(\ref{eq:MHEobjective}); $y_{r, i}$ is the measurement at time $t_i$ from the reactor $r$. The state estimate at the current time, $\left\lbrace x_{\tau, r} | r \in \tilde{\mathcal{R}} \right\rbrace$, can be computed from Eqs.~(\ref{eq:MHEdynamicsConstr})-(\ref{eq:MHEinitConstr}). Note that process noises are not considered in the MHE.

Arrival cost is a crucial aspect of the MHE. The characteristics of the unconsidered past measurement data and previous parameter estimates in $[0, \tau_e]$ are truncated within the arrival cost term, namely $\hat{\theta}$ and $R_0$. Moreover, the positive definite matrix $R_0$ facilitates the stability of the MHE \citep{rawlings2017model}. A standard scheme for the arrival cost design is to use the covariance of the estimation error \citep{elsheikh2021comparative}. The covariance can be approximated with the Cramer-Rao inequality. Prior to the computation, we define $s_r(t_i)=\dfrac{\partial y_r}{\partial \theta} \in \mathbb{R}^{n_y(t_i) \times n_{\theta}}$, the sensitivity matrix of the output of the reactor $r$ with respect to the parameter vector at time $t_i$. The sensitivity matrix propagates through the impulsive dynamics with the following equations:
\begin{equation}
\begin{split}
\dot{s}_r(t) &= \dfrac{\partial f}{\partial x} s_r(t) + \dfrac{\partial f}{\partial \theta}, \quad s_r(0) = \dfrac{\partial x_r(0)}{\partial \theta} \\
s_r(t^+) &= \dfrac{\partial f_d}{\partial x} s_r(t) + \dfrac{\partial f_d}{\partial \theta}
\end{split}
\end{equation}
The covariance for the previous parameter estimate $\hat{\theta}$ can be derived from the MHE objective which has already been utilized in the previous decision times $[0, \tau_e]$ as: 
\begin{equation} \label{eq:MHEArrivalCost}
\begin{split}
F(\tau_e) &=  F(0) + \sum_{r\in \mathcal{\tilde{R}}}\sum_{t_i\in \mathcal{M}_{[0, \tau_e]}} s_r(t_i)^T \Sigma_{t_i}^{-1} s_r(t_i) \\
C_{[0, \tau_e]}(\hat{\theta}) &\approx F(\tau_e)^{-1}
\end{split}
\end{equation}
where $F(\tau_e)$ is referred to as the Fisher information matrix, whose inverse provides a lower bound on the parameter covariance, $ C_{[0, \tau_e]}(\hat{\theta})$. The lower bound of the parameter covariance is then used as $R_0$, i.e., $R_0(\tau_e) = C_{[0, \tau_e]}(\hat{\theta})$ 

Due to a limited amount of at-line measurements, the multi-rate MHE problem is prone to be ill-conditioned. Local parametric sensitivity analysis is utilized to detect and remove non-identifiable parameters that cause such issues. The column rank and singular values are directly related to the rank deficiency of the MHE problem. Meanwhile, the relative scale of the outputs and parameters significantly affects the determination of the identifiable parameters \citep{bae2021multirate}. The scaled sensitivity matrix $\tilde{s}_r(t_i)$ is written as,
\begin{equation}
[\tilde{s}_r(t_i)]_{p, q}  = [s_r(t_i)]_{p, q} \times \dfrac{z_{\theta, q}}{z_{y, p}(t_i)}
\end{equation}
where $z_{y, p}(t_i)$ and $z_{\theta, q}$ represent the scaling factors for the $p^{\text{th}}$ output and $q^{\text{th}}$ parameter. We follow the method suggested in \citep{thompson2009parameter}, where the standard deviations of the output measurement and estimated parameter are employed for the scaling factors. The scaling factors can be expressed in the form of,
\begin{equation}
z_{y, p}(t_i) = \left[diag\left(\Sigma_{t_i}^{1/2} \right) \right]_p, \quad z_{\theta, q} = \left[diag\left(\left( C_{[0, \tau]}(\hat{\theta}) \right)^{-1/2}\right)\right]_q
\end{equation}
where $diag(\cdot)$ is an operator that extracts the diagonal elements of a square matrix.

Now we construct the matrix $S_{\tilde{\mathcal{R}}}\in \mathbb{R}^{\left|\sum_{r\in \tilde{\mathcal{R}}}\sum_{t_i\in \mathcal{M}_{[0, \tau]}}n_y(t_i)\right| \times n_{\theta}}$ with the collection of the scaled sensitivity matrix. The elements of the set $\left\lbrace \tilde{s}_r(t_i) | r\in \tilde{\mathcal{R}}, t_i \in \mathcal{M}_{[0, \tau]} \right\rbrace$ are stacked row-wise. The numerical rank of $S_{\tilde{\mathcal{R}}}$, referred to as $\epsilon$-rank, is the criterion from which the number of the identifiable parameters is determined. It is derived from the singular value spectrum, condition number, and collinearity index of $S_{\tilde{\mathcal{R}}}$ \citep{barz2018adaptive}. Let all singular values $\zeta_i$ be ordered by magnitude starting with $\zeta_1$ being the largest singular value of $S_{\tilde{\mathcal{R}}}$. The numerical $\epsilon$-rank is the maximum number of the singular values $\zeta_i$ that can be added while the sub-condition number $\kappa_{i}=\zeta_1/\zeta_i$ and the sub-collinearity index $\gamma_i=1/\zeta_i$ are below a maximum $\epsilon$-threshold. $\kappa_{max}$ and $\gamma_{max}$ are empirically chosen as $1000$ and $10 - 15$, respectively \citep{lopez2015nonlinear}. The selection of parameters that belong to the identifiable parameter subset is chosen based on a forward selection scheme, namely QRP decomposition. Finally, the original MHE problem in Eq.~(\ref{eq:MHEobjective}) is reduced considering the identifiable subset only.

\subsection{Optimal pulse-feeding strategy with model predictive control}  \label{sec:MPCformulation}
The objective of the MPC is to compute the pulse-feeding strategy (amount of substrate additions in each pulse) that maximizes the biomass and product concentration at the final time (i.e., $t_f$) under the oxygen constraint. According to Fig.~\ref{fig:Colloc_high_resolution}, the feed injections induce instantaneous jumps in the states and fast responses, i.e. drops, in the $DOT$ where $DOT$ easily reaches the limit of the admissible operating region. Meanwhile, the dynamics of biomass growth and product formation are significantly slower. To account for the different rates of the $DOT$ and biomass against the controls, we use specific growth rate $\mu(t)$ instead of biomass concentration $X(t)$ for the objective function.

In addition to the biomass growth, the acetate concentration and feed rate change are controlled. Two important constraints are considered in the optimization problem. First, $DOT$ should always be larger than 20 \% to prevent the cell from being exposed to anaerobic conditions. Second, the feed amount should always be larger than 3 $\mu L$ because zero feed causes cell starvation and pipette action becomes significantly inaccurate when the pulse signal is too low.

The MPC problem is formulated within the time horizon $[\tau, \tau_c]$ for each reactor $r\in \mathcal{R}$ as follows:
\begin{subequations} \label{eq:CultOptProb}
\begin{align}
\min_{u_r} &\quad -Q_0P_r(\tau_c) - Q_1 \int_{\tau}^{\min (\tau_c, t_{ind})}\mu_r(t) dt + Q_2\int_{\tau}^{\min (\tau_c, t_{ind})}A_r(t) dt \nonumber \\
& \quad + \sum_{t \in \mathcal{U}} \| \Delta u_r(t)\|_{Q_3}^2 \\
\text{s.t.}  
& \quad \dot{x}_r(t) = f(x_r(t), \theta_r), \quad t \in [\tau, \tau_c] \setminus \mathcal{U}_{[\tau, \tau_c]} \label{eq:CultOptProb_MKG}\\
& \quad x_r(t^+) = f_d\left(x_r(t), u_r(t)\right), \quad t\in \mathcal{U}_{[\tau, \tau_c]} \label{eq:CultOptProb_MB}\\
& \quad x_{\text{min}} \leq x_r(t) \leq x_{\text{max}}, \quad t \in [\tau, \tau_c]\\
& \quad 3 \mu L \leq u_r(t) \leq 40 \mu L, \quad t \in \mathcal{U}_{[\tau, \tau_c]}  \label{eq:CultOptProb_FeedLB} \\
& \quad DOT_{lb, r} \leq DOT_r(t, \theta_r), \quad t \in \mathcal{U}_{[\tau, \tau_c]} \label{eq:CultOptProb_DOTconst}\\
& \quad b_{r, ind}(t) = \left\lbrace 
\begin{array}{ll}
1 & \text{if} \quad t \geq t_{r, ind} \\
0 & \text{if} \quad t < t_{r, ind}
\end{array} \right. \label{eq:CulOptProb_IndSwitch}\\
& \quad b_{r, ind}(t) \Delta u_r(t) = 0, \quad t\in \mathcal{U}_{[\tau, \tau_c]} \label{eq:CultOptProb_IndConstFeed} \\
& \quad x_r(0)=x_{0, r},
\end{align}
\end{subequations}
where $Q_0$, $Q_1$, $Q_2$, and $Q_3$ are positive constant weighting factors for the final time product $P_r$, cumulative biomass growth rate $\mu$, cumulative acetate $A_r$, and feed rate changes $\Delta u_r$, respectively; subscript $r$ represents that the variable is for the reactor $r\in \mathcal{R}$; $x_{min}$ and $x_{max}$ are the lower and upper bounds for the state variables; $DOT_{lb, r}$ represents the DOT lower bound of reactor $r$. Eq.~(\ref{eq:CulOptProb_IndSwitch}) describes that the induction switch $b_{r, ind}$ is turned on after the induction time $t_{r, ind}$. According to the constraint Eq.~(\ref{eq:CultOptProb_IndConstFeed}), the pulse feeds become constant during the induction phase. The optimization problem considers up to the prediction horizon $\tau_c$. Full horizon is not recommended, because both the objective and constraint function are model (parameter) dependent. 

Objective and constraint functions which are related to the pulse feed are only defined in the discrete-time set $\mathcal{U}_{[\tau, \tau_c]}$. Here, the oxygen constraint Eq.~(\ref{eq:CultOptProb_DOTconst}) is also defined in $\mathcal{U}_{[\tau, \tau_c]}$, because the minimum value of DOT within a single pulse is always located at which pulse-feed is given. Eqs.~(\ref{eq:CultOptProb}) are reformulated to the standard nonlinear programming problem (NLP) by discretizing the differential equations. Full discretization method is implemented to capture the dynamics at $[\tau, \tau_c] \setminus \mathcal{U}_{[\tau, \tau_c]}$. Orthogonal collocation with the Gauss-Radau method is employed \citep{biegler2010nonlinear}.

\subsection{Closed-loop integration of MHE and MPC}
\label{sec:MHEMPCintegration}
In the real-time closed-loop implementation, MHE and MPC are implemented $\forall t \in \mathcal{U}$ in a receding horizon manner. Utilizing the measurement data, parameters and states are estimated by the MHE with the estimation horizon. The MPC acquires the current state estimate from the MHE and computes the optimal pulse-feed trajectory up to the prediction horizon. In addition, at the interface between MPC and the actuation, the time for the computation time for MHE and MPC and actuation and measurement time in the robotic facility should be considered. To account for the overall delay, MPC regards the starting time as the current time added by the time offset, $t_{offset}$. The filtered initial state for the MPC should be computed based on that time offset, i.e., $\hat{x}_{t + t_{offset}}$.

\begin{figure}
\centering
\includegraphics[width=\linewidth]{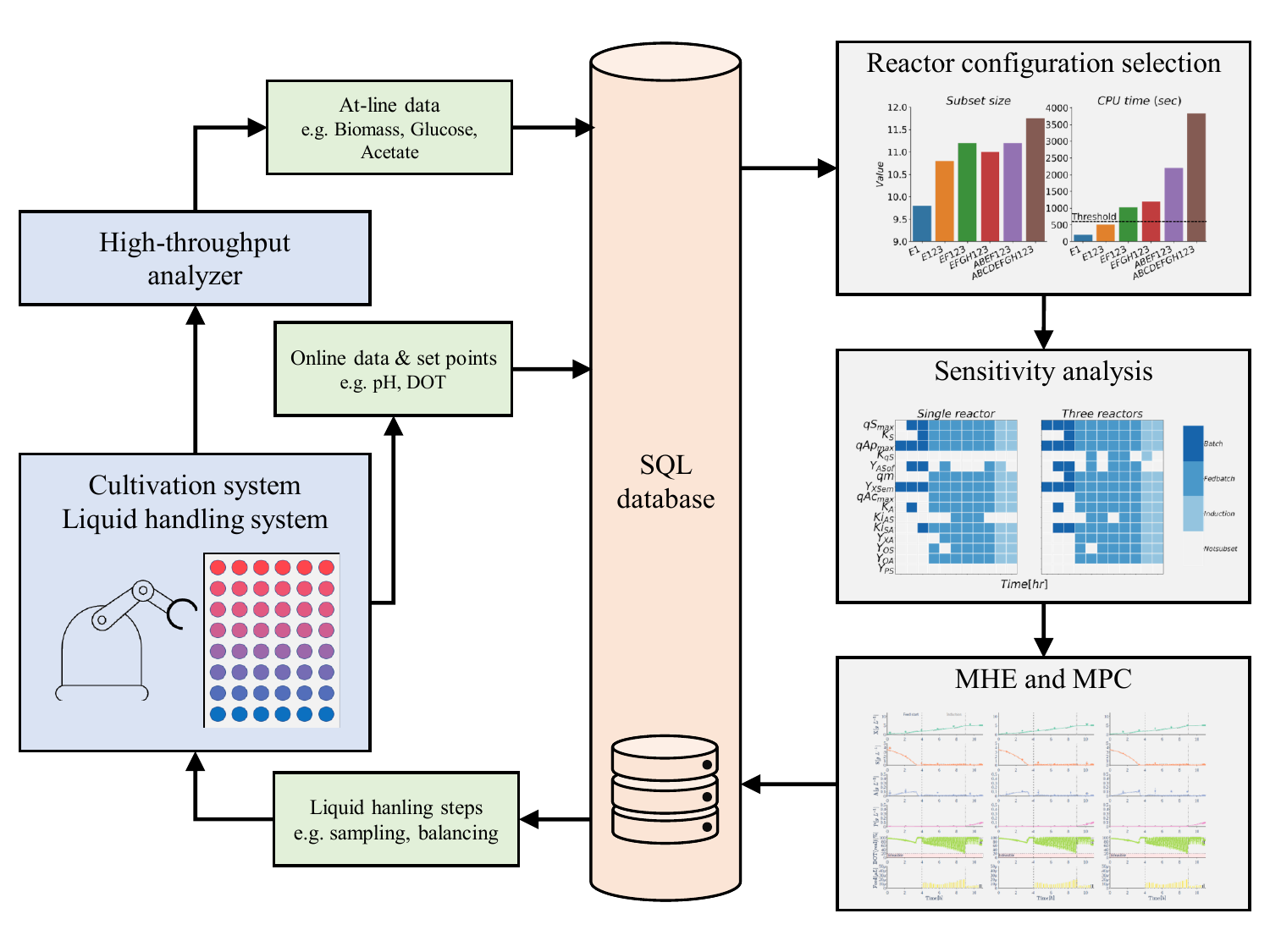} 
\caption{Schematic diagram of the automated and closed-loop cultivation.}
\label{fig:MHEMPCProcedure}
\end{figure}

As introduced in Sec.~\ref{sec:Intro}, the framework provides automated and closed-loop cultivation that 1) performs the pre-scheduled pulse-feeding strategy, 2) measures the experimental data, 3) communicates via database, 4) validates and fits the macro-kinetic growth model to the data, and 5) re-optimizes the pulse-feeding strategy. Figure~\ref{fig:MHEMPCProcedure} illustrates the procedure of the proposed framework.

\section{\textit{In silico} results}
\subsection{Determination of the MHE configuration and identifiability analysis}
The \textit{in silico} cultivation for 24 mini-bioreactors is performed in parallel, with the proposed model-based methods. To generate the \textit{in silico} data, we add random uniform noise with 20 \% of their scale to the growth parameters in Table~\ref{tb:MKG_global_params} for each row and add additional uniform noise with 5 \% of their scale to the replicate in each column. For the reactor-dependent parameters in Table~\ref{tb:MKG_local_params}, 20 \% uniform noises are added for all bioreactors. Moreover, Gaussian noise with 5 \% variance to the scales is added to the measurement. The prediction horizon and estimation horizon were 180 and 300 min, respectively. A standard setting of the collocation method for the continuous control, the third degree of the orthogonal polynomial and one element, failed to model the fluctuation from the pulse input. Thus, a denser collocation method with the third degree of the orthogonal polynomial with three elements was used.

Do-mpc software was used to formulate the macro-kinetic growth model and the MPC \citep{lucia2017rapid}. Do-mpc is a toolbox developed based on CasADi, a software framework for nonlinear optimization and optimal control \citep{Andersson2019}. The modularized representation of the mathematical model as a graphical structure allows for the fast calculation of the differentiation, hence enabling efficient ODE integration and gradient-based optimization, such as IPOPT \citep{wachter2006implementation}.

Before the state and parameter estimation, it has to be decided which reactors to consider in a single MHE optimization. We propose two criteria for the decision, identifiable growth parameter subset size and computational time of the MHE. Only the growth parameters (see Table~\ref{tb:MKG_global_params}) are accounted for in the parameter subset. The proposed MHE and MPC were solved for the entire horizon and three criteria were averaged. There are six candidate configurations denoted as `E1', `E123', `EF123', `EFGH123', `ABEF123', and `ABCDEFGH123', whose alphabet part and numeric part represent the row and column of the reactors, respectively. According to the cultivation condition of each row in Table~\ref{tb:MBR8condns}, `E1', `E123', `EF1', and `EFGH123' incorporate the measurement data generated from the same parameter set, whereas `ABEF123', and `ABCDEFGH123' accommodate the measurement data generated from different parameter sets. 

\begin{figure}
\centering
\includegraphics[width=\linewidth]{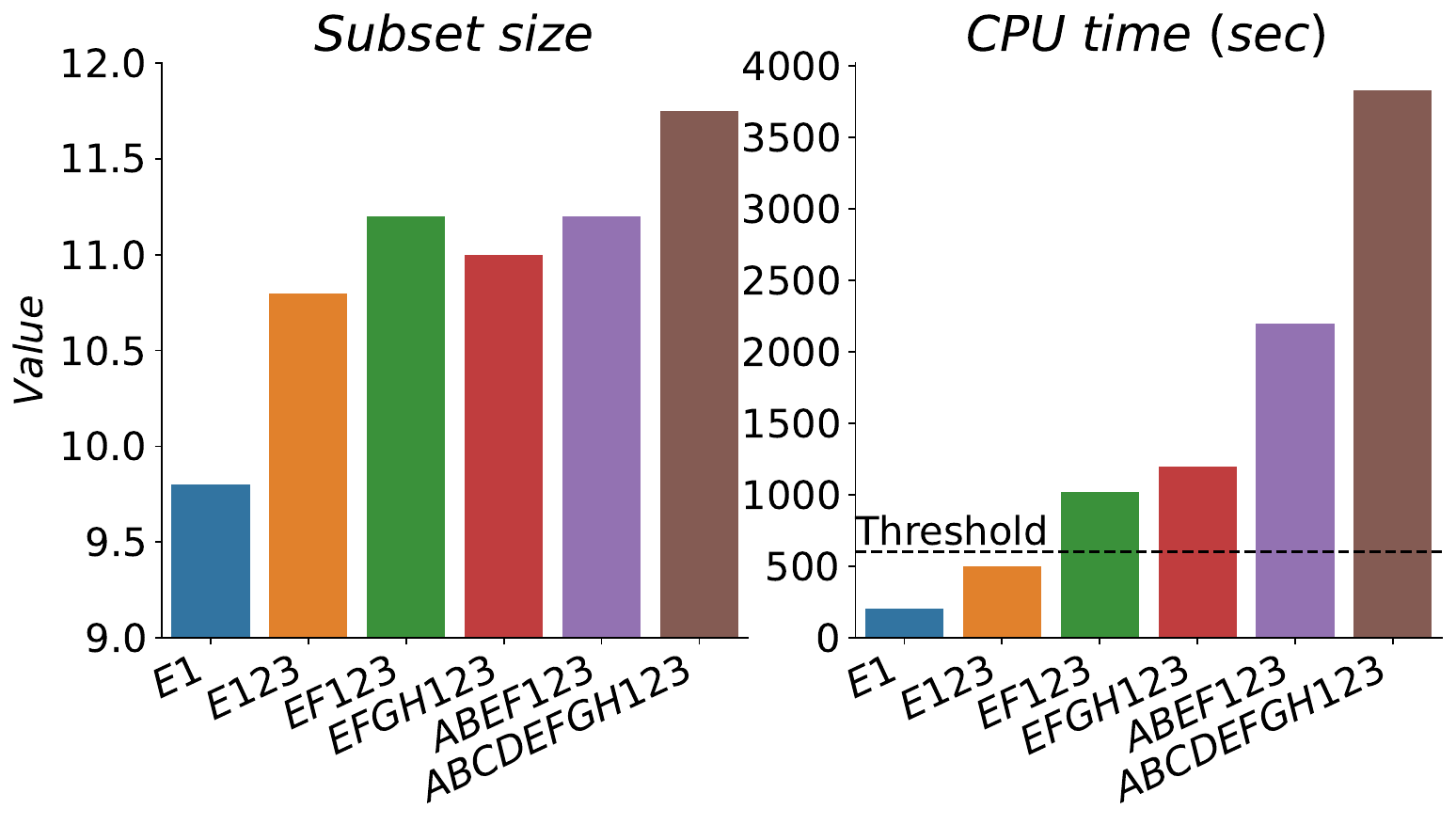} 
\caption{Sizes of identifiable growth parameter subset and CPU times of the MHE under six different optimization configurations. Average values over ten runs are plotted.}
\label{fig:pe_plot}
\end{figure}

Figure~\ref{fig:pe_plot} shows the average values of subset size and CPU time over ten runs of the MHE under six different optimization configurations. Regarding the subset size, except for two extreme cases `E1' and `ABCDEFGH123', the other four configurations show a similar number of identifiable growth parameters. Computational time is proportional to the number of reactors included in a single MHE optimization. A good configuration should possess a large number of identifiable parameter subset sizes and a shorter computational time than the decision interval, of 10 min. Based on the observation, we conclude that `E123', likewise, `A123', `B123', $\ldots$, and `H123', are the most suitable configurations that satisfy such requirements.

\begin{figure}
\centering
\includegraphics[width=\linewidth]{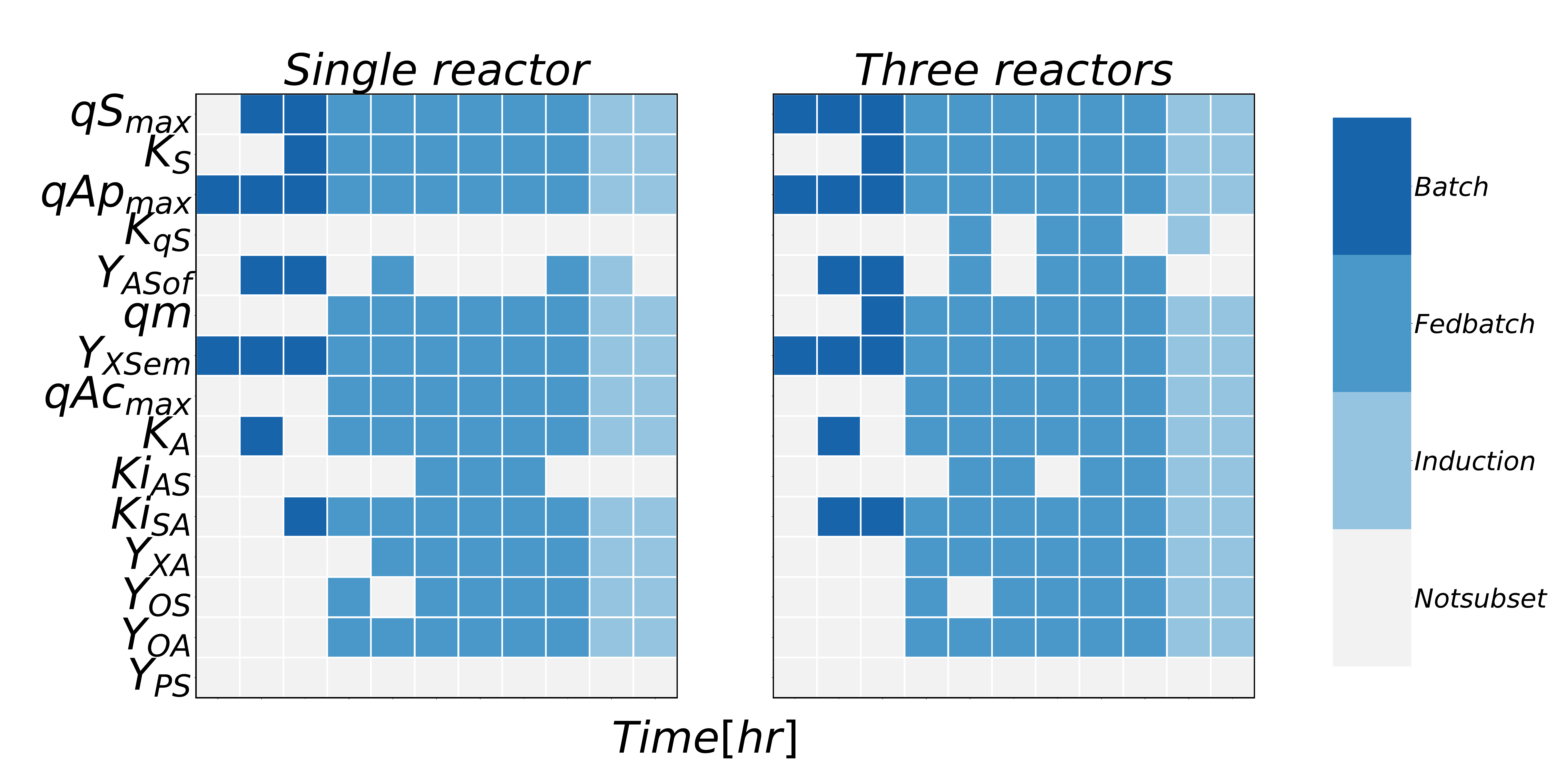} 
\caption{Identifiable parameters evaluated from incorporating a single reactor (left) and triplicate reactors (right). The color bar indicates the cultivation phase at which the identifiability analysis is performed.}
\label{fig:param_subset}
\end{figure}

A detailed result of the identifiable analysis under two configurations of using a single reactor (i.e., `E1') and three reactors (i.e., `E123') is depicted in Fig.~\ref{fig:param_subset}. It highlights how identifiable parameter subset is progressively varying throughout the cultivation time, which is divided into three phases, batch, fed-batch, and induction. In general, the subset size is higher when a larger number of measurement data is employed. Particularly, the parameter $K_{qS}$ is completely non-identifiable in the first case, while it becomes identifiable in the second case. Nonetheless, the parameter $Y_{PS}$ is non-identifiable in both cases, since the amount of non-zero product measurement is not enough for the $Y_{PS}$-column of the sensitivity matrix $S_{\tilde{\mathcal{R}}}$ to exceed the $\epsilon$-threshold. The result also demonstrates that the subset size dramatically grows in the fed-batch phase. The fluctuating pattern of the states and specific rates originating from the pulse feed leads to the increasing number of non-zero elements of the sensitivity matrix $S_{\tilde{\mathcal{R}}}$ and its column-rank. It can be seen that the pulse feed gives a persistent excitation to the system and is therefore advantageous in terms of identifiability.

\subsection{MPC results}

This section presents the MPC results based on the moving horizon state and parameter estimation results. As explained in Section~\ref{sec:MBR}, the cultivation is divided into three phases: the batch phase, feeding phase, and induction phase. the batch phase ends when the substrate is depleted. We observe the depletion by measuring the sudden increase in the DOT and starting the feeding phase. In the feeding phase, the substrate is either fed in a pulse using an exponential strategy or an MPC strategy. A standard approach for the pulse-feed calculation is the exponential feed given as follows \citep{sawatzki2018accelerated, hans2020automated}: 
\begin{equation}
\Delta v = \left( \dfrac{\mu_{set}}{ Y_{XS}} + q_m \right) \dfrac{X_0 V_0}{S_f}  e^{\mu_{set} \cdot t}
\end{equation}
where $\mu_{set} \ (h^{-1})$ represents the desired growth rate; $X_0 \ (g/L)$ and $V_0 \ (L)$ represent the biomass concentration and volume at the end of the batch phase, respectively. Finally, an inducer is added to produce the recombinant protein and a constant feed is applied.

\begin{figure}
\centering
\includegraphics[width=\linewidth]{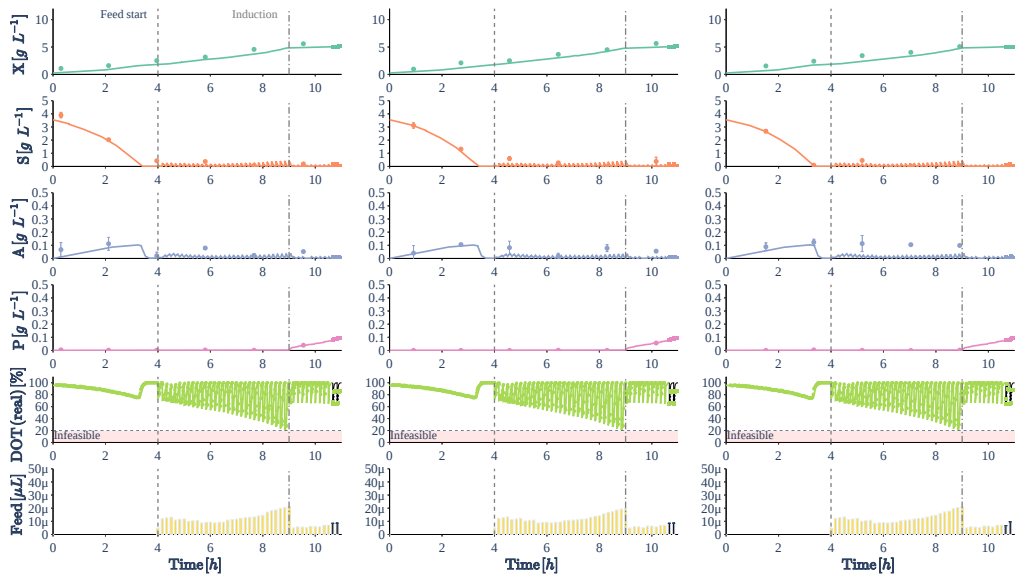}
\caption{State and input trajectories for the three replicates of the reactor E under 11
hours of cultivation. Symbols of y axis represent following; $X$: biomass, $S$: Substrate, $A$: Acetate, $P$: Product, $DOT$: Dissolved oxygen tension. Dotted and solid lines represent the measurement and simulated trajectories, respectively.}
\label{fig:E_triplicate}
\end{figure}

Figure~\ref{fig:E_triplicate} shows the 
state (i.e., biomass, substrate, acetic acid, DOT measurement, product, and DOT actual) and input (i.e., pulse-feed) trajectories for the three replicates of reactor E. The reactor E is cultivated under the condition of initial biomass 3.55 ($g/L$) and glucose 0.25 ($g/L$), DOT bound 20\%, and strain I. According to the decision criteria in Fig.~\ref{fig:pe_plot}, three reactors of `E123' are considered in a single MHE optimization. After the 4 hours of batch phase, the substrate initially given is depleted and DOT increases to the saturated concentration. MPC starts to compute the pulse feed from the fed-batch phase. The actual DOT reaches to the bound of 20\% at the end of the fed-batch phase. At hour 9, the induction agent is implemented and the product starts to be produced under the near constant pulse feed. 

\begin{figure}
\centering
\includegraphics[width=\linewidth]{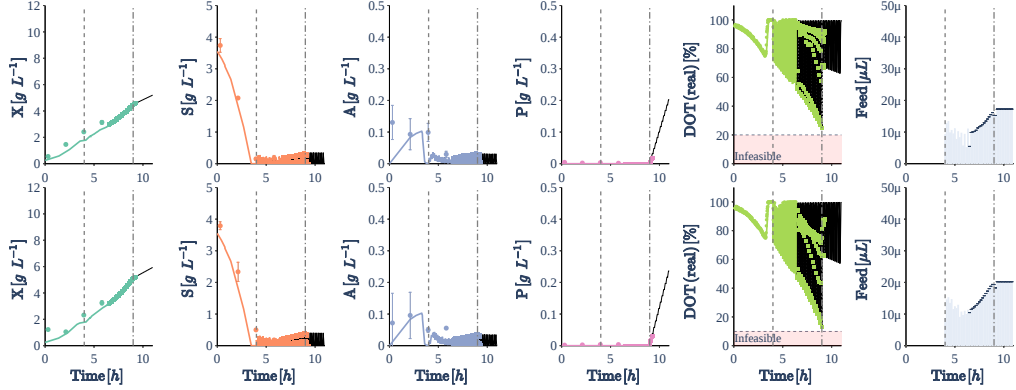} 
\caption{Comparison between the reactors with two different DOT constraints, i.e., reactor E (top) and reactor F (bottom). Graphs show the snapshot at hour 6.33. The colored solid line represents the \textit{in silico} state trajectory, the dot represents the measurement, the black line represents the predicted state trajectory, and the dotted line represents collocation points from the optimization problem of MPC.}
\label{fig:EF_diff_DOT}
\end{figure}

We now demonstrate the results of the reactors F, G, and A whose conditions differ from the reactor E in the DOT constraint, initial state condition, and strain respectively. Figure~\ref{fig:EF_diff_DOT} illustrates the MPC results of the reactor rows E and F, which differ in the DOT lower bound (i.e., 20\% and 10\%, respectively). Reactor F has a lower DOT bound than E, so a larger amount of pulse feeds are allowed. The average feed amounts for reactors E and F within the prediction horizon are 14.8 $\mu L$ and 16.9 $\mu L$, respectively. Note that the feed values beyond the prediction horizon are arbitrary, so they are not included in the analysis.

\begin{figure}
\centering
\includegraphics[width=\linewidth]{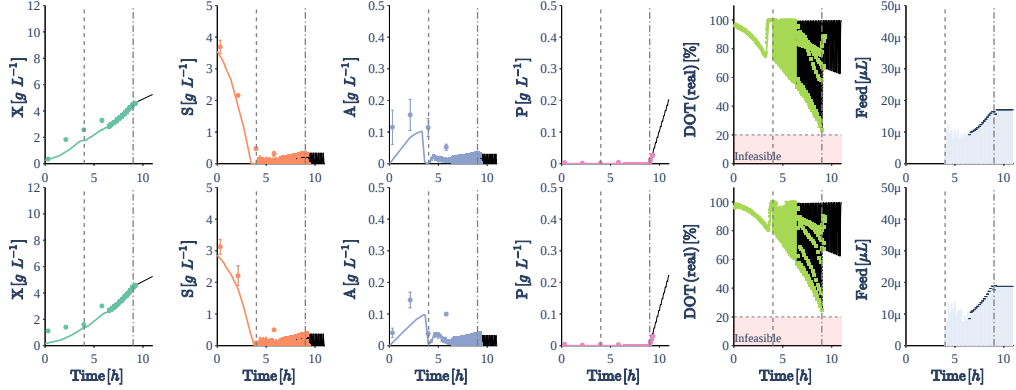} 
\caption{Comparison between the reactors with two different initial biomass and glucose conditions, i.e., reactor E (top) and reactor G (bottom). Graphs show the snapshot at hour 6.33. The colored solid line represents the \textit{in silico} state trajectory, the dot represents the measurement, and the black line represents the predicted state trajectory. The dotted line represents collocation points of the state trajectory, and the dotted line represents collocation points from the optimization problem of MPC.}
\label{fig:EG_diff_init}
\end{figure}

Similarly, the reactor row E and G, which have different initial biomass and substrate conditions, are compared in Fig.~\ref{fig:EG_diff_init}. Since reactor G has lower initial concentrations (i.e., initial biomass 2.86 ($g/L$) and glucose 0.16 ($g/L$)), the amount of biomass at the batch-end is lower than that of reactor E. The smaller amount of biomass that exists, the smaller amount of oxygen consumed. Consequently, reactor G allows for a higher pulse-feed amount for the DOT to reach the boundary. The average feed amounts for reactors E and G within the prediction horizon are 14.9 $\mu L$ and 16.3 $\mu L$, respectively.

\begin{figure}
\centering
\includegraphics[width=\linewidth]{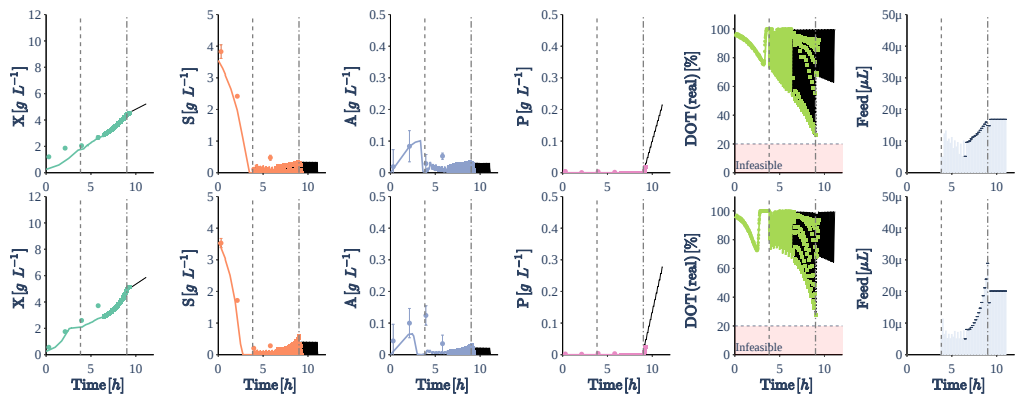} 
\caption{Comparison between the reactors with two different strains, i.e., reactor E (top) and reactor A (bottom). Graphs show the snapshot at hour 6.33. The colored solid line represents the \textit{in silico} state trajectory, the dot represents the measurement, the black line represents the predicted state trajectory, and the dotted line represents collocation points from the optimization problem of MPC.}
\label{fig:EA_diff_param}
\end{figure}

Figure~\ref{fig:EA_diff_param} shows the MPC results of reactor rows E and A, whose strains are II and I (see Table~\ref{tb:MKG_global_params}), respectively. The reactor E has smaller $\mu$ and $q_S$, and larger $q_A$. The trend is shown in Fig.~\ref{fig:EA_diff_param}, in the sense that the batch ends faster in reactor A, with smaller production of acetic acid. The result also indicates that the MPC is able to compute the feasible pulse-feed strategy, in which the complex parametric relation is taken into account.


\begin{figure}
\centering
\includegraphics[width=\linewidth]{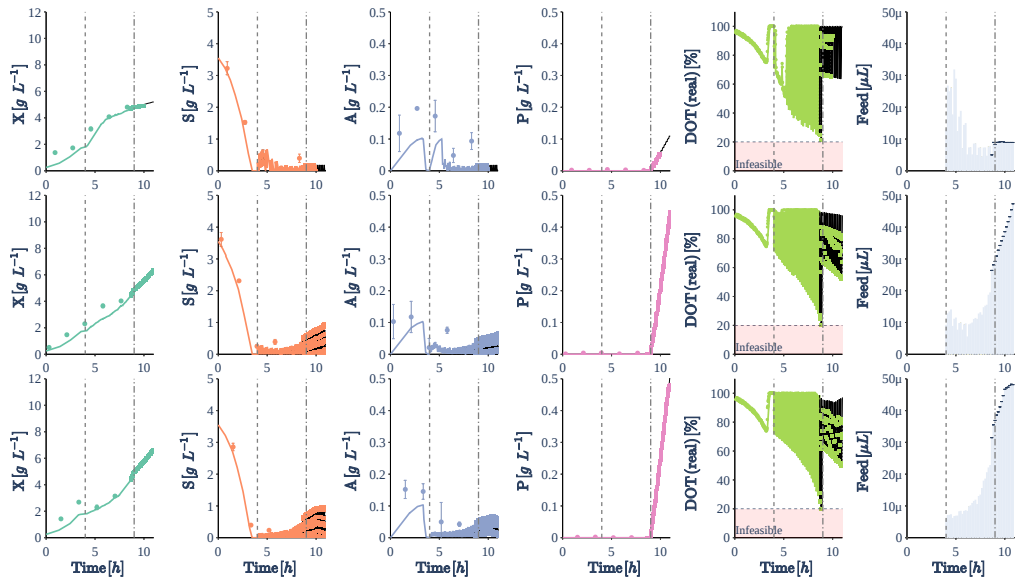} 
\caption{Comparison between the reactors with different prediction horizons, i.e., 90 min (top), 180 min (middle), and 270 min (bottom). Graphs show the snapshot at hour 9. The colored solid line represents the \textit{in silico} state trajectory, the dot represents the measurement, the black line represents the predicted state trajectory, and the dotted line represents collocation points from the optimization problem of MPC.}
\label{fig:Different_horizon}
\end{figure}

The effect of the prediction horizon on the MPC result is investigated. Three reactor results with different prediction horizons are illustrated in Fig.~\ref{fig:Different_horizon}. The biomass concentrations of the three reactors at the end of the cultivation are 4.78, 6.32, and 6.59 g/L, respectively, and the product concentrations of the three reactors at the end of the cultivation are 0.17, 0.45, and 0.49 g/L, respectively. When a short prediction horizon is used, a larger amount of pulse feeds are given in the early feeding phase. The cell grows earlier, and the DOT reaches the lower bound earlier than the optimal feeding strategy. Because the cell grows exponentially, it is advantageous to feed more in the later feeding stage. The long-term feeding strategy can only be developed when MPC uses a long prediction horizon. Still, there exists a disadvantage to using a full prediction horizon. When the model-plant mismatch is severe, estimation results get closer too late to be corrected.

\section{Concluding remarks}
In this study, we develop a macro-kinetic growth model-based MHE and MPC for the high-throughput experiment with 24 parallel mini-bioreactors. Three crucial aspects of the system are addressed. First, the pulse feed is formulated as an impulsive control system. We perform the analysis of the objective and constraint function and implemented the full-discretization method for solving the MPC. Second, multi-rate MHE is formulated. The MHE method utilizes parameter sensitivity as an important measure. The scaled Fisher information matrix is used as an arriving cost of the MHE. In addition, the dynamic sensitivity matrix is used to capture the identifiable parameter subset. Third, the reactor configuration for the MHE is determined based on the size of the identifiable parameter subset and computational time. Through an offline computation, E123' is selected as the optimal configuration among six reactor configurations, (`E1', `E123', `EF123', `EFGH123', `ABEF123', and `ABCDEFGH123'). Because `E123' has enough subset size compared to the larger configurations and it requires less computation time than the threshold, 600 sec. 

We perform an online \textit{in silico} study for reactors with 8 different cultivation conditions with 3 replicates each. The effects of the DOT lower bound, strain type, and initial biomass/substrate concentrations on the model-based MHE and MPC are discussed. We show that the proposed framework is capable of providing feed strategies adaptively to different conditions. When the DOT lower bound is changed from 20\% to 10\%, the pulse feeds are increased by 14\%. Similarly, when the initial substrate/biomass concentrations are changed from 3.55/0.25 g/L to 2.86/0.16 g/L, respectively, the pulse feeds are increased by 9.4\%. The MPC can also compute feasible and consistent pulse feed amounts under another strain. In addition, we analyzed the effect of the prediction horizon on cultivation. Compare to the shortest prediction horizon of 90 min, the MPCs with prediction horizons of 180 and 270 min lead to increased biomass concentrations of 32\% and 37\%, respectively, and increased product concentrations of 165\% and 288\%, respectively. 

The major drawback of the model-based methods is that the performance depends strongly on the accuracy of the macro-kinetic growth model. In fact, due to the complexity of the cell metabolism, the kinetic parameters vary between different cultivation conditions \citep{anane2019output}. The replicates present different dynamics due to cellular adaptation. Even in the single bioreactor, the parameters are strongly correlated to the feed trajectory and its frequency \citep{anane2019modelling}. These uncertainties make the experimental validation challenging as demonstrated in the experimental results \citep{krausch2022htbd}. Therefore, the future research direction is to implement the robust controller which computes more conservative feeds depending on the uncertainty.

\section*{Acknowledgment}
This work was supported by the German Federal Ministry of Education and Research through the Program International Future Labs for Artificial Intelligence (Grant number 01DD20002A).
\appendix

\section{Detailed description of macro-kinetic \textit{E. coli} growth model} \label{Apx:MKGmodel}

The cell grows from the consequence of the metabolic pathway in which glucose (substrate) and oxygen are utilized. Glucose is distributed to the physiological activities in the cell such as oxidative and overflow routes. Energy for growth, production, and cell maintenance is obtained from the oxidative route associated with oxygen usage. Glucose is further converted to acetate in the overflow mechanism, which is described by the concept of acetate cycling. The overall intracellular kinetics are modeled by a set of algebraic equations \citep{anane2017modelling}. 

Substrate uptake rate $q_S$ has Monod kinetics and is inhibited non-competitively by the acetate concentration as
\begin{equation}
q_S = q_{S, max}\dfrac{S}{S + K_S} \cdot \dfrac{K_{i,SA}}{K_{i,SA} + A}  
\end{equation}
where $q_{S, max}$ ($g/(g\cdot h)$) is the maximum specific substrate uptake rate; $K_S$ ($g/L$) denote the affinity constant for substrate consumption; $K_{i,SA}$ ($g/L$) is the inhibition constant of substrate uptake by acetate. Specific cell growth rate $\mu$ is given as
\begin{equation}
\mu = q_{S,an} + q_{Ac}Y_{XA}
\end{equation}
where $q_{S,an}$ ($g/(g\cdot h)$) represents the substrate utilization rate for anabolism (see Eq.~(\ref{eq:MKG_qSan})); $q_{Ac}$ ($g/(g\cdot h)$) is the specific acetate consumption rate (see Eq.~(\ref{eq:MKG_qAc})); $Y_{XA}$ ($g/g$) is the yield coefficient of biomass over acetate. The glucose partitioning is described as
\begin{equation}
q_S = q_{S,ox} + q_{S,of}
\end{equation}
where $q_{S,ox}$ and $q_{S,of}$ ($g/(g\cdot h)$) represent the specific substrate uptake rate for overflow and oxidative routes, respectively. $q_{S,of}$ is obtained by
\begin{equation}
q_{S,of} = \dfrac{q_{Ap}}{Y_{AS,of}}
\end{equation}
where $q_{Ap}$ is the specific acetate production rate defined in Eq.~(\ref{eq:MKG_qAp}); $Y_{AS,of}$ ($g/(g\cdot h)$) is the yield coefficient of acetate over the substrate in overflow metabolism. The oxidative pathway comprises anabolic $q_{S,an}$ and energetic $q_{S,en}$ cell activities, and energetic activity is further divided into cell maintenance $q_{m}$ and product formation $q_{S,ox, P}$. As a result, the following relationship can be obtained:
\begin{align} 
    q_{S,an} &= \left( q_{S,ox} - q_{m} - q_{S,ox, P} \right)Y_{XS,em}    \label{eq:MKG_qSan}\\
    q_{S,en} &= q_{S,ox} - q_{m} - q_{S,ox, P} - q_{S,an}
\end{align}
where $Y_{XS,em}$ ($g/g$) is the yield coefficient of biomass over the substrate from exclusive maintenance. Product is generated from the oxidative flow according to the following relationship:
\begin{equation} \label{eq:MKG_qSoxP}
q_{S,ox, P} = d_{S,ox, P}\cdot q_{S,ox} \cdot s_{ind}, 
\end{equation}
where $d_{S,ox, P}$ is the distribution ratio; $s_{ind}$ is the switch indicating the induction agent is given. 

Acetate is produced and consumed with the rates $q_{Ap}$ and $q_{Ac}$ ($g/(g\cdot h)$), respectively, i.e., $q_A = q_{Ap} - q_{Ac}$. Acetate production is limited by the substrate uptake rate $q_S$ with the Monod-type kinetics as
\begin{equation} \label{eq:MKG_qAp}
q_{Ap} = q_{Ap, max} \dfrac{q_S}{q_S + K_{qS}}
\end{equation}
where $q_{Ap, max}$ ($g/(g\cdot h)$) is the maximum specific intracellular uptake rate for acetate flux; $K_{qS}$ ($g/(g\cdot h)$) represents the affinity constant to intracellular substrate flux. Acetate consumption is modeled by Monod kinetics with non-competitive inhibition as
\begin{equation} \label{eq:MKG_qAc}
q_{Ac} = q_{Ac, max}\dfrac{A}{A + K_{A}}\cdot\dfrac{K_{i,AS}}{K_{i,AS} + q_S}
\end{equation}
where $q_{Ac, max}$ ($g/(g\cdot h)$) is the maximum specific acetate consumption rate; $K_A$ ($g/L$) is affinity constant for acetate consumption; $K_{i,AS}$ ($g/L$) is the inhibition constant of acetate uptake by the substrate.

Product formation rate $q_P$ is proportional to the specific product formation rate from substrate flux $q_{S,ox, P}$ as
\begin{equation}
q_P = q_{S,ox, P}Y_{PS}
\end{equation}
where $Y_{PS}$ ($g/g$) is the yield coefficient of product over substrate.

Dissolved oxygen tension (DOT) is typically modeled by the differential equation as
\begin{equation}
\dfrac{dDOT}{dt} = k_{la}(DOT^* - DOT) - q_OXH
\end{equation}
where $k_{la}$ ($h^{-1}$) denotes the volumetric oxygen transfer coefficient; $DOT^*$ ($\%$) denotes the saturation concentration of DOT; $q_O$ ($g/(g\cdot h)$) denotes the specific oxygen uptake rate (see Eq.~(\ref{eq:MKG_qO})); $H$ ($mol/(m^3 \cdot Pa)$) denotes the Henry constant. Considering that DOT presents the fast dynamics \citep{duan2020model}, we use the reduced form expressed in the algebraic equation as
\begin{equation} \label{eq:MKG_DOTalg}
DOT = DOT^* - \dfrac{q_OXH}{k_{la}}
\end{equation}
Specific oxygen uptake rate $q_O$ is obtained by
\begin{equation} \label{eq:MKG_qO}
q_O = (q_{S,en} + q_{m})Y_{OS} + q_{Ac}Y_{OA}
\end{equation}
where $Y_{OS}$ and $Y_{OA}$ ($g/g$) are the yield coefficients of the oxygen over biomass and acetate, respectively. 

\section{MPC results of eight experimental designs}

\begin{figure}
\centering
\includegraphics[width=\linewidth]{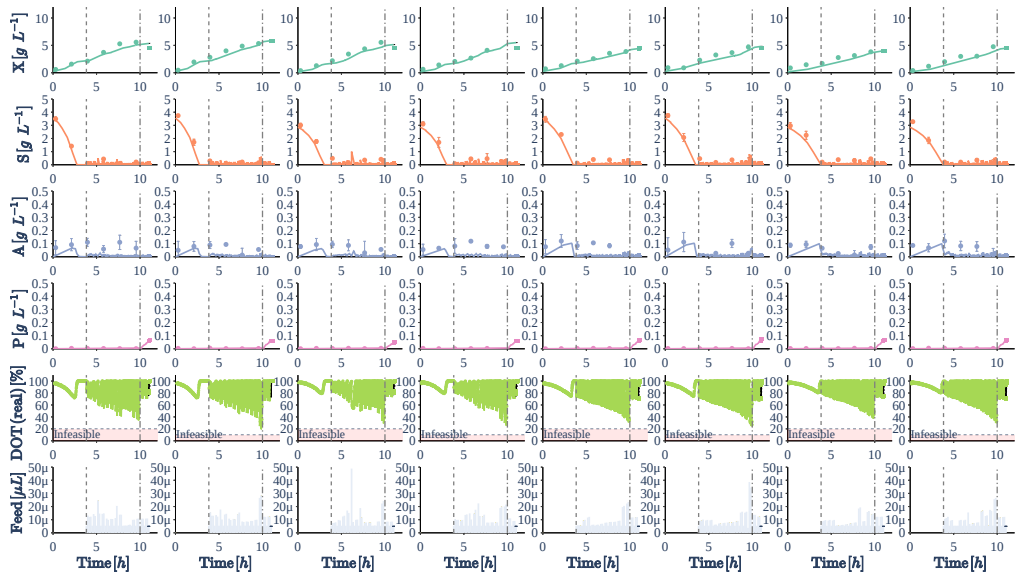} 
\caption{MPC results of eight experimental designs. Symbols of y axis represent following; $X$: biomass, $S$: Substrate, $A$: Acetate, $P$: Product, $DOT$: Dissolved oxygen tension.}
\label{fig:Allreactors}
\end{figure}

Figure~\ref{fig:Allreactors} shows the MPC results of all eight experimental designs. One reactor per triplicate is plotted.

\bibliographystyle{elsarticle-harv} 
\bibliography{cas-refs}

\begin{thebibliography}{70}
\expandafter\ifx\csname natexlab\endcsname\relax\def\natexlab#1{#1}\fi
\providecommand{\url}[1]{\texttt{#1}}
\providecommand{\href}[2]{#2}
\providecommand{\path}[1]{#1}
\providecommand{\DOIprefix}{doi:}
\providecommand{\ArXivprefix}{arXiv:}
\providecommand{\URLprefix}{URL: }
\providecommand{\Pubmedprefix}{pmid:}
\providecommand{\doi}[1]{\href{http://dx.doi.org/#1}{\path{#1}}}
\providecommand{\Pubmed}[1]{\href{pmid:#1}{\path{#1}}}
\providecommand{\bibinfo}[2]{#2}
\ifx\xfnm\relax \def\xfnm[#1]{\unskip,\space#1}\fi
\bibitem[{Abdollahi and Dubljevic(2012)}]{abdollahi2012lipid}
\bibinfo{author}{Abdollahi, J.}, \bibinfo{author}{Dubljevic, S.},
  \bibinfo{year}{2012}.
\newblock \bibinfo{title}{{Lipid production optimization and optimal control of
  heterotrophic microalgae fed-batch bioreactor}}.
\newblock \bibinfo{journal}{Chemical engineering science} \bibinfo{volume}{84},
  \bibinfo{pages}{619--627}.
\bibitem[{Alexander et~al.(2020)Alexander, Campani, Dinh and
  Lima}]{alexander2020challenges}
\bibinfo{author}{Alexander, R.}, \bibinfo{author}{Campani, G.},
  \bibinfo{author}{Dinh, S.}, \bibinfo{author}{Lima, F.V.},
  \bibinfo{year}{2020}.
\newblock \bibinfo{title}{Challenges and opportunities on nonlinear state
  estimation of chemical and biochemical processes}.
\newblock \bibinfo{journal}{Processes} \bibinfo{volume}{8},
  \bibinfo{pages}{1462}.
\bibitem[{Anane et~al.(2019a)Anane, Barz, Sin, Gernaey, Neubauer, Bournazou
  et~al.}]{anane2019output}
\bibinfo{author}{Anane, E.}, \bibinfo{author}{Barz, T.}, \bibinfo{author}{Sin,
  G.}, \bibinfo{author}{Gernaey, K.V.}, \bibinfo{author}{Neubauer, P.},
  \bibinfo{author}{Bournazou, M.N.C.}, et~al., \bibinfo{year}{2019}a.
\newblock \bibinfo{title}{{Output uncertainty of dynamic growth models: effect
  of uncertain parameter estimates on model reliability}}.
\newblock \bibinfo{journal}{Biochemical Engineering Journal}
  \bibinfo{volume}{150}, \bibinfo{pages}{107247}.
\bibitem[{Anane et~al.(2017)Anane, Lopez~C{\'a}rdenas, Neubauer and
  Cruz~Bournazou}]{anane2017modelling}
\bibinfo{author}{Anane, E.}, \bibinfo{author}{Lopez~C{\'a}rdenas, D.C.},
  \bibinfo{author}{Neubauer, P.}, \bibinfo{author}{Cruz~Bournazou, M.N.},
  \bibinfo{year}{2017}.
\newblock \bibinfo{title}{{Modelling overflow metabolism in Escherichia coli by
  acetate cycling}}.
\newblock \bibinfo{journal}{Biochemical engineering journal}
  \bibinfo{volume}{125}, \bibinfo{pages}{23--30}.
\bibitem[{Anane et~al.(2019b)Anane, Sawatzki, Neubauer and
  Cruz-Bournazou}]{anane2019modelling}
\bibinfo{author}{Anane, E.}, \bibinfo{author}{Sawatzki, A.},
  \bibinfo{author}{Neubauer, P.}, \bibinfo{author}{Cruz-Bournazou, M.N.},
  \bibinfo{year}{2019}b.
\newblock \bibinfo{title}{{Modelling concentration gradients in fed-batch
  cultivations of E. coli - towards the flexible design of scale-down
  experiments}}.
\newblock \bibinfo{journal}{Journal of Chemical Technology \& Biotechnology}
  \bibinfo{volume}{94}, \bibinfo{pages}{516--526}.
\bibitem[{Andersson et~al.(2019)Andersson, Gillis, Horn, Rawlings and
  Diehl}]{Andersson2019}
\bibinfo{author}{Andersson, J.A.E.}, \bibinfo{author}{Gillis, J.},
  \bibinfo{author}{Horn, G.}, \bibinfo{author}{Rawlings, J.B.},
  \bibinfo{author}{Diehl, M.}, \bibinfo{year}{2019}.
\newblock \bibinfo{title}{{{CasADi} -- {A} software framework for nonlinear
  optimization and optimal control}}.
\newblock \bibinfo{journal}{Mathematical Programming Computation}
  \bibinfo{volume}{11}, \bibinfo{pages}{1--36}.
\newblock \DOIprefix\doi{10.1007/s12532-018-0139-4}.
\bibitem[{Ashoori et~al.(2009)Ashoori, Moshiri, Khaki-Sedigh and
  Bakhtiari}]{ashoori2009optimal}
\bibinfo{author}{Ashoori, A.}, \bibinfo{author}{Moshiri, B.},
  \bibinfo{author}{Khaki-Sedigh, A.}, \bibinfo{author}{Bakhtiari, M.R.},
  \bibinfo{year}{2009}.
\newblock \bibinfo{title}{{Optimal control of a nonlinear fed-batch
  fermentation process using model predictive approach}}.
\newblock \bibinfo{journal}{Journal of Process Control} \bibinfo{volume}{19},
  \bibinfo{pages}{1162--1173}.
\bibitem[{Bae et~al.(2021)Bae, Kim and Lee}]{bae2021multirate}
\bibinfo{author}{Bae, J.}, \bibinfo{author}{Kim, Y.}, \bibinfo{author}{Lee,
  J.M.}, \bibinfo{year}{2021}.
\newblock \bibinfo{title}{Multirate moving horizon estimation combined with
  parameter subset selection}.
\newblock \bibinfo{journal}{Computers \& Chemical Engineering}
  \bibinfo{volume}{147}, \bibinfo{pages}{107253}.
\bibitem[{Barz et~al.(2016)Barz, Bournazou, K{\"o}rkel, Walter
  et~al.}]{barz2016real}
\bibinfo{author}{Barz, T.}, \bibinfo{author}{Bournazou, M.N.C.},
  \bibinfo{author}{K{\"o}rkel, S.}, \bibinfo{author}{Walter, S.F.}, et~al.,
  \bibinfo{year}{2016}.
\newblock \bibinfo{title}{{Real-time adaptive input design for the
  determination of competitive adsorption isotherms in liquid chromatography}}.
\newblock \bibinfo{journal}{Computers \& Chemical Engineering}
  \bibinfo{volume}{94}, \bibinfo{pages}{104--116}.
\bibitem[{Barz et~al.(2018)Barz, Sommer, Wilms, Neubauer and
  Bournazou}]{barz2018adaptive}
\bibinfo{author}{Barz, T.}, \bibinfo{author}{Sommer, A.},
  \bibinfo{author}{Wilms, T.}, \bibinfo{author}{Neubauer, P.},
  \bibinfo{author}{Bournazou, M.N.C.}, \bibinfo{year}{2018}.
\newblock \bibinfo{title}{Adaptive optimal operation of a parallel robotic
  liquid handling station}.
\newblock \bibinfo{journal}{IFAC-PapersOnLine} \bibinfo{volume}{51},
  \bibinfo{pages}{765--770}.
\bibitem[{Biegler(2010)}]{biegler2010nonlinear}
\bibinfo{author}{Biegler, L.T.}, \bibinfo{year}{2010}.
\newblock \bibinfo{title}{{Nonlinear programming: concepts, algorithms, and
  applications to chemical processes}}.
\newblock \bibinfo{publisher}{SIAM}.
\bibitem[{Bunzel et~al.(2018)Bunzel, Garrabou, Pott and
  Hilvert}]{bunzel2018speeding}
\bibinfo{author}{Bunzel, H.A.}, \bibinfo{author}{Garrabou, X.},
  \bibinfo{author}{Pott, M.}, \bibinfo{author}{Hilvert, D.},
  \bibinfo{year}{2018}.
\newblock \bibinfo{title}{{Speeding up enzyme discovery and engineering with
  ultrahigh-throughput methods}}.
\newblock \bibinfo{journal}{Current opinion in structural biology}
  \bibinfo{volume}{48}, \bibinfo{pages}{149--156}.
\bibitem[{Cruz~Bournazou et~al.(2017)Cruz~Bournazou, Barz, Nickel,
  Lopez~C{\'a}rdenas, Glauche, Knepper and Neubauer}]{cruz2017online}
\bibinfo{author}{Cruz~Bournazou, M.N.}, \bibinfo{author}{Barz, T.},
  \bibinfo{author}{Nickel, D.}, \bibinfo{author}{Lopez~C{\'a}rdenas, D.},
  \bibinfo{author}{Glauche, F.}, \bibinfo{author}{Knepper, A.},
  \bibinfo{author}{Neubauer, P.}, \bibinfo{year}{2017}.
\newblock \bibinfo{title}{{Online optimal experimental re-design in robotic
  parallel fed-batch cultivation facilities}}.
\newblock \bibinfo{journal}{Biotechnology and bioengineering}
  \bibinfo{volume}{114}, \bibinfo{pages}{610--619}.
\bibitem[{Delvigne et~al.(2009)Delvigne, Boxus, Ingels and
  Thonart}]{delvigne2009bioreactor}
\bibinfo{author}{Delvigne, F.}, \bibinfo{author}{Boxus, M.},
  \bibinfo{author}{Ingels, S.}, \bibinfo{author}{Thonart, P.},
  \bibinfo{year}{2009}.
\newblock \bibinfo{title}{{Bioreactor mixing efficiency modulates the activity
  of a prpoS:: GFP reporter gene in E. coli}}.
\newblock \bibinfo{journal}{Microbial cell factories} \bibinfo{volume}{8},
  \bibinfo{pages}{1--17}.
\bibitem[{Diehl et~al.(2006)Diehl, K{\"u}hl, Bock and
  Schl{\"o}der}]{diehl2006schnelle}
\bibinfo{author}{Diehl, M.}, \bibinfo{author}{K{\"u}hl, P.},
  \bibinfo{author}{Bock, H.G.}, \bibinfo{author}{Schl{\"o}der, J.P.},
  \bibinfo{year}{2006}.
\newblock \bibinfo{title}{Schnelle algorithmen f{\"u}r die zustands-und
  parametersch{\"a}tzung auf bewegten horizonten (fast algorithms for state and
  parameter estimation on moving horizons)}.
\newblock \bibinfo{journal}{Automatisierungstechnik} \bibinfo{volume}{54},
  \bibinfo{pages}{602--613}.
\bibitem[{Duan et~al.(2020)Duan, Wilms, Neubauer, Kravaris and
  Cruz~Bournazou}]{duan2020model}
\bibinfo{author}{Duan, Z.}, \bibinfo{author}{Wilms, T.},
  \bibinfo{author}{Neubauer, P.}, \bibinfo{author}{Kravaris, C.},
  \bibinfo{author}{Cruz~Bournazou, M.N.}, \bibinfo{year}{2020}.
\newblock \bibinfo{title}{Model reduction of aerobic bioprocess models for
  efficient simulation}.
\newblock \bibinfo{journal}{Chemical Engineering Science}
  \bibinfo{volume}{217}, \bibinfo{pages}{115512}.
\bibitem[{Elsheikh et~al.(2021)Elsheikh, Hille, Tatulea-Codrean and
  Kr{\"a}mer}]{elsheikh2021comparative}
\bibinfo{author}{Elsheikh, M.}, \bibinfo{author}{Hille, R.},
  \bibinfo{author}{Tatulea-Codrean, A.}, \bibinfo{author}{Kr{\"a}mer, S.},
  \bibinfo{year}{2021}.
\newblock \bibinfo{title}{A comparative review of multi-rate moving horizon
  estimation schemes for bioprocess applications}.
\newblock \bibinfo{journal}{Computers \& Chemical Engineering} ,
  \bibinfo{pages}{107219}.
\bibitem[{Faust et~al.(2014)Faust, Janzen, Bendig, R{\"o}mer, Kaufmann and
  Weuster-Botz}]{faust2014feeding}
\bibinfo{author}{Faust, G.}, \bibinfo{author}{Janzen, N.H.},
  \bibinfo{author}{Bendig, C.}, \bibinfo{author}{R{\"o}mer, L.},
  \bibinfo{author}{Kaufmann, K.}, \bibinfo{author}{Weuster-Botz, D.},
  \bibinfo{year}{2014}.
\newblock \bibinfo{title}{Feeding strategies enhance high cell density
  cultivation and protein expression in milliliter scale bioreactors}.
\newblock \bibinfo{journal}{Biotechnology journal} \bibinfo{volume}{9},
  \bibinfo{pages}{1293--1303}.
\bibitem[{Fiedler et~al.(2020)Fiedler, Baumbach, B{\"o}rner and
  Lucia}]{fiedler2020probabilistic}
\bibinfo{author}{Fiedler, F.}, \bibinfo{author}{Baumbach, D.},
  \bibinfo{author}{B{\"o}rner, A.}, \bibinfo{author}{Lucia, S.},
  \bibinfo{year}{2020}.
\newblock \bibinfo{title}{{A Probabilistic Moving Horizon Estimation Framework
  Applied to the Visual-Inertial Sensor Fusion Problem}}, in:
  \bibinfo{booktitle}{2020 European Control Conference (ECC)},
  \bibinfo{organization}{IEEE}. pp. \bibinfo{pages}{1009--1016}.
\bibitem[{Fink et~al.(2021)Fink, Cserjan-Puschmann, Reinisch and
  Striedner}]{fink2021high}
\bibinfo{author}{Fink, M.}, \bibinfo{author}{Cserjan-Puschmann, M.},
  \bibinfo{author}{Reinisch, D.}, \bibinfo{author}{Striedner, G.},
  \bibinfo{year}{2021}.
\newblock \bibinfo{title}{{High-throughput microbioreactor provides a capable
  tool for early stage bioprocess development}}.
\newblock \bibinfo{journal}{Scientific reports} \bibinfo{volume}{11},
  \bibinfo{pages}{1--10}.
\bibitem[{Freitas et~al.(2017)Freitas, Olivo and
  Andrade}]{freitas2017optimization}
\bibinfo{author}{Freitas, H.F.S.d.}, \bibinfo{author}{Olivo, J.E.},
  \bibinfo{author}{Andrade, C.M.G.}, \bibinfo{year}{2017}.
\newblock \bibinfo{title}{{Optimization of bioethanol in silico production
  process in a fed-batch bioreactor using non-linear model predictive control
  and evolutionary computation techniques}}.
\newblock \bibinfo{journal}{Energies} \bibinfo{volume}{10},
  \bibinfo{pages}{1763}.
\bibitem[{Gomes et~al.(2015)Gomes, Chopda and Rathore}]{gomes2015integrating}
\bibinfo{author}{Gomes, J.}, \bibinfo{author}{Chopda, V.R.},
  \bibinfo{author}{Rathore, A.S.}, \bibinfo{year}{2015}.
\newblock \bibinfo{title}{Integrating systems analysis and control for
  implementing process analytical technology in bioprocess development}.
\newblock \bibinfo{journal}{Journal of Chemical Technology \& Biotechnology}
  \bibinfo{volume}{90}, \bibinfo{pages}{583--589}.
\bibitem[{Haby et~al.(2019)Haby, Hans, Anane, Sawatzki, Krausch, Neubauer and
  Cruz~Bournazou}]{haby2019integrated}
\bibinfo{author}{Haby, B.}, \bibinfo{author}{Hans, S.}, \bibinfo{author}{Anane,
  E.}, \bibinfo{author}{Sawatzki, A.}, \bibinfo{author}{Krausch, N.},
  \bibinfo{author}{Neubauer, P.}, \bibinfo{author}{Cruz~Bournazou, M.N.},
  \bibinfo{year}{2019}.
\newblock \bibinfo{title}{{Integrated robotic mini bioreactor platform for
  automated, parallel microbial cultivation with online data handling and
  process control}}.
\newblock \bibinfo{journal}{SLAS TECHNOLOGY: Translating Life Sciences
  Innovation} \bibinfo{volume}{24}, \bibinfo{pages}{569--582}.
\bibitem[{Hans et~al.(2020a)Hans, Haby, Krausch, Barz, Neubauer and
  Cruz~Bournazou}]{hans2020automated}
\bibinfo{author}{Hans, S.}, \bibinfo{author}{Haby, B.},
  \bibinfo{author}{Krausch, N.}, \bibinfo{author}{Barz, T.},
  \bibinfo{author}{Neubauer, P.}, \bibinfo{author}{Cruz~Bournazou, M.N.},
  \bibinfo{year}{2020}a.
\newblock \bibinfo{title}{{Automated Conditional Screening of Multiple
  Escherichia coli Strains in Parallel Adaptive Fed-Batch Cultivations}}.
\newblock \bibinfo{journal}{Bioengineering} \bibinfo{volume}{7},
  \bibinfo{pages}{145}.
\bibitem[{Hans et~al.(2020b)Hans, Ulmer, Narayanan, Brautaset, Krausch,
  Neubauer, Sch{\"a}ffl, Sokolov and Cruz~Bournazou}]{hans2020monitoring}
\bibinfo{author}{Hans, S.}, \bibinfo{author}{Ulmer, C.},
  \bibinfo{author}{Narayanan, H.}, \bibinfo{author}{Brautaset, T.},
  \bibinfo{author}{Krausch, N.}, \bibinfo{author}{Neubauer, P.},
  \bibinfo{author}{Sch{\"a}ffl, I.}, \bibinfo{author}{Sokolov, M.},
  \bibinfo{author}{Cruz~Bournazou, M.N.}, \bibinfo{year}{2020}b.
\newblock \bibinfo{title}{{Monitoring Parallel Robotic Cultivations with Online
  Multivariate Analysis}}.
\newblock \bibinfo{journal}{Processes} \bibinfo{volume}{8},
  \bibinfo{pages}{582}.
\bibitem[{Hemmerich et~al.(2018)Hemmerich, Noack, Wiechert and
  Oldiges}]{hemmerich2018microbioreactor}
\bibinfo{author}{Hemmerich, J.}, \bibinfo{author}{Noack, S.},
  \bibinfo{author}{Wiechert, W.}, \bibinfo{author}{Oldiges, M.},
  \bibinfo{year}{2018}.
\newblock \bibinfo{title}{Microbioreactor systems for accelerated bioprocess
  development}.
\newblock \bibinfo{journal}{Biotechnology journal} \bibinfo{volume}{13},
  \bibinfo{pages}{1700141}.
\bibitem[{Hemmerich et~al.(2021)Hemmerich, Tenhaef, Wiechert and
  Noack}]{hemmerich2021pyfoomb}
\bibinfo{author}{Hemmerich, J.}, \bibinfo{author}{Tenhaef, N.},
  \bibinfo{author}{Wiechert, W.}, \bibinfo{author}{Noack, S.},
  \bibinfo{year}{2021}.
\newblock \bibinfo{title}{{pyFOOMB: Python framework for object oriented
  modeling of bioprocesses}}.
\newblock \bibinfo{journal}{Engineering in Life Sciences} \bibinfo{volume}{21},
  \bibinfo{pages}{242--257}.
\bibitem[{Herwig et~al.(2021)Herwig, P{\"o}rtner and
  M{\"o}ller}]{herwig2021digital}
\bibinfo{author}{Herwig, C.}, \bibinfo{author}{P{\"o}rtner, R.},
  \bibinfo{author}{M{\"o}ller, J.}, \bibinfo{year}{2021}.
\newblock \bibinfo{title}{{Digital Twins: Applications to the Design and
  Optimization of Bioprocesses}}. volume \bibinfo{volume}{177}.
\newblock \bibinfo{publisher}{Springer Nature}.
\bibitem[{Janzen et~al.(2019)Janzen, Striedner, Jarmer, Voigtmann, Abad and
  Reinisch}]{janzen2019implementation}
\bibinfo{author}{Janzen, N.H.}, \bibinfo{author}{Striedner, G.},
  \bibinfo{author}{Jarmer, J.}, \bibinfo{author}{Voigtmann, M.},
  \bibinfo{author}{Abad, S.}, \bibinfo{author}{Reinisch, D.},
  \bibinfo{year}{2019}.
\newblock \bibinfo{title}{Implementation of a fully automated microbial
  cultivation platform for strain and process screening}.
\newblock \bibinfo{journal}{Biotechnology journal} \bibinfo{volume}{14},
  \bibinfo{pages}{1800625}.
\bibitem[{Jouned et~al.(2022)Jouned, Kager, Herwig and
  Tilman}]{jouned2022event}
\bibinfo{author}{Jouned, M.A.}, \bibinfo{author}{Kager, J.},
  \bibinfo{author}{Herwig, C.}, \bibinfo{author}{Tilman, B.},
  \bibinfo{year}{2022}.
\newblock \bibinfo{title}{{Event driven modelling for the accurate
  identification of metabolic switches in fed-batch culture of S. cerevisiae}}.
\newblock \bibinfo{journal}{Biochemical Engineering Journal} ,
  \bibinfo{pages}{108345}.
\bibitem[{Kim et~al.(2021a)Kim, Krausch, Aizpuru, Barz, Lucia, Mart{\'\i}nez,
  Neubauer and Bournazou}]{kim2021oed}
\bibinfo{author}{Kim, J.W.}, \bibinfo{author}{Krausch, N.},
  \bibinfo{author}{Aizpuru, J.}, \bibinfo{author}{Barz, T.},
  \bibinfo{author}{Lucia, S.}, \bibinfo{author}{Mart{\'\i}nez, E.C.},
  \bibinfo{author}{Neubauer, P.}, \bibinfo{author}{Bournazou, M.N.C.},
  \bibinfo{year}{2021}a.
\newblock \bibinfo{title}{Model predictive control guided with optimal
  experimental design for pulse-based parallel cultivation}.
\newblock \bibinfo{journal}{arXiv preprint arXiv:2112.10548} .
\bibitem[{Kim et~al.(2021b)Kim, Park, Oh and Lee}]{kim2021model}
\bibinfo{author}{Kim, J.W.}, \bibinfo{author}{Park, B.J.}, \bibinfo{author}{Oh,
  T.H.}, \bibinfo{author}{Lee, J.M.}, \bibinfo{year}{2021}b.
\newblock \bibinfo{title}{{Model-based reinforcement learning and predictive
  control for two-stage optimal control of fed-batch bioreactor}}.
\newblock \bibinfo{journal}{Computers \& Chemical Engineering}
  \bibinfo{volume}{154}, \bibinfo{pages}{107465}.
\bibitem[{Kr{\"a}mer and Gesthuisen(2005)}]{kramer2005multirate}
\bibinfo{author}{Kr{\"a}mer, S.}, \bibinfo{author}{Gesthuisen, R.},
  \bibinfo{year}{2005}.
\newblock \bibinfo{title}{Multirate state estimation using moving horizon
  estimation}.
\newblock \bibinfo{journal}{IFAC Proceedings Volumes} \bibinfo{volume}{38},
  \bibinfo{pages}{1--6}.
\bibitem[{Kramer et~al.(2005)Kramer, Gesthuisen and Engell}]{kramer2005fixed}
\bibinfo{author}{Kramer, S.}, \bibinfo{author}{Gesthuisen, R.},
  \bibinfo{author}{Engell, S.}, \bibinfo{year}{2005}.
\newblock \bibinfo{title}{Fixed structure multirate state estimation}, in:
  \bibinfo{booktitle}{Proceedings of the 2005, American Control Conference,
  2005.}, \bibinfo{organization}{IEEE}. pp. \bibinfo{pages}{4613--4618}.
\bibitem[{Krausch et~al.(2022)Krausch, Kim, Barz, Lucia, Gro{\ss}, Huber,
  Schiller, Neubauer and Cruz~Bournazou}]{krausch2022htbd}
\bibinfo{author}{Krausch, N.}, \bibinfo{author}{Kim, J.W.},
  \bibinfo{author}{Barz, T.}, \bibinfo{author}{Lucia, S.},
  \bibinfo{author}{Gro{\ss}, S.}, \bibinfo{author}{Huber, M.C.},
  \bibinfo{author}{Schiller, S.M.}, \bibinfo{author}{Neubauer, P.},
  \bibinfo{author}{Cruz~Bournazou, M.N.}, \bibinfo{year}{2022}.
\newblock \bibinfo{title}{High-throughput screening of optimal process
  conditions using model predictive control}.
\newblock \bibinfo{journal}{Biotechnology and Bioengineering}
  \bibinfo{volume}{119}, \bibinfo{pages}{3584--3595}.
\bibitem[{Kusterer et~al.(2008)Kusterer, Krause, Kaufmann, Arnold and
  Weuster-Botz}]{kusterer2008fully}
\bibinfo{author}{Kusterer, A.}, \bibinfo{author}{Krause, C.},
  \bibinfo{author}{Kaufmann, K.}, \bibinfo{author}{Arnold, M.},
  \bibinfo{author}{Weuster-Botz, D.}, \bibinfo{year}{2008}.
\newblock \bibinfo{title}{{Fully automated single-use stirred-tank bioreactors
  for parallel microbial cultivations}}.
\newblock \bibinfo{journal}{Bioprocess and biosystems engineering}
  \bibinfo{volume}{31}, \bibinfo{pages}{207--215}.
\bibitem[{Leavell et~al.(2020)Leavell, Singh and
  Kaufmann-Malaga}]{leavell2020high}
\bibinfo{author}{Leavell, M.D.}, \bibinfo{author}{Singh, A.H.},
  \bibinfo{author}{Kaufmann-Malaga, B.B.}, \bibinfo{year}{2020}.
\newblock \bibinfo{title}{{High-throughput screening for improved microbial
  cell factories, perspective and promise}}.
\newblock \bibinfo{journal}{Current opinion in biotechnology}
  \bibinfo{volume}{62}, \bibinfo{pages}{22--28}.
\bibitem[{Liu et~al.(2016)Liu, Zhang, Chen and Yu}]{liu2016moving}
\bibinfo{author}{Liu, A.}, \bibinfo{author}{Zhang, W.A.},
  \bibinfo{author}{Chen, M.Z.}, \bibinfo{author}{Yu, L.}, \bibinfo{year}{2016}.
\newblock \bibinfo{title}{Moving horizon estimation for mobile robots with
  multirate sampling}.
\newblock \bibinfo{journal}{IEEE Transactions on Industrial Electronics}
  \bibinfo{volume}{64}, \bibinfo{pages}{1457--1467}.
\bibitem[{L\'opez~C\'ardenas et~al.(2015)L\'opez~C\'ardenas, Barz, K\"orkel and
  Wozny}]{lopez2015nonlinear}
\bibinfo{author}{L\'opez~C\'ardenas, D.C.}, \bibinfo{author}{Barz, T.},
  \bibinfo{author}{K\"orkel, S.}, \bibinfo{author}{Wozny, G.},
  \bibinfo{year}{2015}.
\newblock \bibinfo{title}{{Nonlinear ill-posed problem analysis in model-based
  parameter estimation and experimental design}}.
\newblock \bibinfo{journal}{Computers \& Chemical Engineering}
  \bibinfo{volume}{77}, \bibinfo{pages}{24--42}.
\bibitem[{L{\'o}pez-Negrete and Biegler(2012)}]{lopez2012moving}
\bibinfo{author}{L{\'o}pez-Negrete, R.}, \bibinfo{author}{Biegler, L.T.},
  \bibinfo{year}{2012}.
\newblock \bibinfo{title}{{A moving horizon estimator for processes with
  multi-rate measurements: A nonlinear programming sensitivity approach}}.
\newblock \bibinfo{journal}{Journal of Process Control} \bibinfo{volume}{22},
  \bibinfo{pages}{677--688}.
\bibitem[{Lucia et~al.(2017a)Lucia, Carius and Findeisen}]{lucia2017adaptive}
\bibinfo{author}{Lucia, S.}, \bibinfo{author}{Carius, L.},
  \bibinfo{author}{Findeisen, R.}, \bibinfo{year}{2017}a.
\newblock \bibinfo{title}{Adaptive nonlinear predictive control and estimation
  of microaerobic processes}.
\newblock \bibinfo{journal}{IFAC-PapersOnLine} \bibinfo{volume}{50},
  \bibinfo{pages}{12635--12640}.
\bibitem[{Lucia et~al.(2017b)Lucia, T{\u{a}}tulea-Codrean, Schoppmeyer and
  Engell}]{lucia2017rapid}
\bibinfo{author}{Lucia, S.}, \bibinfo{author}{T{\u{a}}tulea-Codrean, A.},
  \bibinfo{author}{Schoppmeyer, C.}, \bibinfo{author}{Engell, S.},
  \bibinfo{year}{2017}b.
\newblock \bibinfo{title}{{Rapid development of modular and sustainable
  nonlinear model predictive control solutions}}.
\newblock \bibinfo{journal}{Control Engineering Practice} \bibinfo{volume}{60},
  \bibinfo{pages}{51--62}.
\bibitem[{Luna and Mart{\'\i}nez(2017)}]{luna2017iterative}
\bibinfo{author}{Luna, M.F.}, \bibinfo{author}{Mart{\'\i}nez, E.C.},
  \bibinfo{year}{2017}.
\newblock \bibinfo{title}{Iterative modeling and optimization of biomass
  production using experimental feedback}.
\newblock \bibinfo{journal}{Computers \& Chemical Engineering}
  \bibinfo{volume}{104}, \bibinfo{pages}{151--163}.
\bibitem[{Markana et~al.(2018)Markana, Padhiyar and
  Moudgalya}]{markana2018multi}
\bibinfo{author}{Markana, A.}, \bibinfo{author}{Padhiyar, N.},
  \bibinfo{author}{Moudgalya, K.}, \bibinfo{year}{2018}.
\newblock \bibinfo{title}{{Multi-criterion control of a bioprocess in fed-batch
  reactor using EKF based economic model predictive control}}.
\newblock \bibinfo{journal}{Chemical Engineering Research and Design}
  \bibinfo{volume}{136}, \bibinfo{pages}{282--294}.
\bibitem[{Mart{\'\i}nez et~al.(2013)Mart{\'\i}nez, Cristaldi and
  Grau}]{martinez2013dynamic}
\bibinfo{author}{Mart{\'\i}nez, E.C.}, \bibinfo{author}{Cristaldi, M.D.},
  \bibinfo{author}{Grau, R.J.}, \bibinfo{year}{2013}.
\newblock \bibinfo{title}{{Dynamic optimization of bioreactors using
  probabilistic tendency models and Bayesian active learning}}.
\newblock \bibinfo{journal}{Computers \& Chemical Engineering}
  \bibinfo{volume}{49}, \bibinfo{pages}{37--49}.
\bibitem[{Narayanan et~al.(2020)Narayanan, Luna, von Stosch, Cruz~Bournazou,
  Polotti, Morbidelli, Butt{\'e} and Sokolov}]{narayanan2020bioprocessing}
\bibinfo{author}{Narayanan, H.}, \bibinfo{author}{Luna, M.F.},
  \bibinfo{author}{von Stosch, M.}, \bibinfo{author}{Cruz~Bournazou, M.N.},
  \bibinfo{author}{Polotti, G.}, \bibinfo{author}{Morbidelli, M.},
  \bibinfo{author}{Butt{\'e}, A.}, \bibinfo{author}{Sokolov, M.},
  \bibinfo{year}{2020}.
\newblock \bibinfo{title}{{Bioprocessing in the digital age: the role of
  process models}}.
\newblock \bibinfo{journal}{Biotechnology journal} \bibinfo{volume}{15},
  \bibinfo{pages}{1900172}.
\bibitem[{Neubauer et~al.(2013)Neubauer, Cruz, Glauche, Junne, Knepper and
  Raven}]{neubauer2013consistent}
\bibinfo{author}{Neubauer, P.}, \bibinfo{author}{Cruz, N.},
  \bibinfo{author}{Glauche, F.}, \bibinfo{author}{Junne, S.},
  \bibinfo{author}{Knepper, A.}, \bibinfo{author}{Raven, M.},
  \bibinfo{year}{2013}.
\newblock \bibinfo{title}{Consistent development of bioprocesses from
  microliter cultures to the industrial scale}.
\newblock \bibinfo{journal}{Engineering in Life Sciences} \bibinfo{volume}{13},
  \bibinfo{pages}{224--238}.
\bibitem[{Neubauer and Junne(2016)}]{neubauer2016scale}
\bibinfo{author}{Neubauer, P.}, \bibinfo{author}{Junne, S.},
  \bibinfo{year}{2016}.
\newblock \bibinfo{title}{{Scale-up and scale-down methodologies for
  bioreactors}}.
\newblock \bibinfo{publisher}{Chichester: John Wiley \& Sons}.
\bibitem[{P{\v{c}}olka and {\v{C}}elikovsk{\`y}(2016)}]{pvcolka2016algorithms}
\bibinfo{author}{P{\v{c}}olka, M.}, \bibinfo{author}{{\v{C}}elikovsk{\`y}, S.},
  \bibinfo{year}{2016}.
\newblock \bibinfo{title}{{Algorithms for nonlinear predictive control
  maximizing penicillin production efficiency}}, in: \bibinfo{booktitle}{2016
  American Control Conference (ACC)}, \bibinfo{organization}{IEEE}. pp.
  \bibinfo{pages}{3527--3532}.
\bibitem[{Petsagkourakis et~al.(2020)Petsagkourakis, Sandoval, Bradford, Zhang
  and del Rio-Chanona}]{petsagkourakis2020reinforcement}
\bibinfo{author}{Petsagkourakis, P.}, \bibinfo{author}{Sandoval, I.O.},
  \bibinfo{author}{Bradford, E.}, \bibinfo{author}{Zhang, D.},
  \bibinfo{author}{del Rio-Chanona, E.A.}, \bibinfo{year}{2020}.
\newblock \bibinfo{title}{Reinforcement learning for batch bioprocess
  optimization}.
\newblock \bibinfo{journal}{Computers \& Chemical Engineering}
  \bibinfo{volume}{133}, \bibinfo{pages}{106649}.
\bibitem[{Puskeiler et~al.(2005)Puskeiler, Kaufmann and
  Weuster-Botz}]{puskeiler2005development}
\bibinfo{author}{Puskeiler, R.}, \bibinfo{author}{Kaufmann, K.},
  \bibinfo{author}{Weuster-Botz, D.}, \bibinfo{year}{2005}.
\newblock \bibinfo{title}{{Development, parallelization, and automation of a
  gas-inducing milliliter-scale bioreactor for high-throughput bioprocess
  design (HTBD)}}.
\newblock \bibinfo{journal}{Biotechnology and bioengineering}
  \bibinfo{volume}{89}, \bibinfo{pages}{512--523}.
\bibitem[{Qu and Hahn(2009)}]{qu2009computation}
\bibinfo{author}{Qu, C.C.}, \bibinfo{author}{Hahn, J.}, \bibinfo{year}{2009}.
\newblock \bibinfo{title}{{Computation of arrival cost for moving horizon
  estimation via unscented Kalman filtering}}.
\newblock \bibinfo{journal}{Journal of Process Control} \bibinfo{volume}{19},
  \bibinfo{pages}{358--363}.
\bibitem[{Raftery et~al.(2017)Raftery, DeSessa and Karim}]{raftery2017economic}
\bibinfo{author}{Raftery, J.P.}, \bibinfo{author}{DeSessa, M.R.},
  \bibinfo{author}{Karim, M.N.}, \bibinfo{year}{2017}.
\newblock \bibinfo{title}{{Economic improvement of continuous pharmaceutical
  production via the optimal control of a multifeed bioreactor}}.
\newblock \bibinfo{journal}{Biotechnology progress} \bibinfo{volume}{33},
  \bibinfo{pages}{902--912}.
\bibitem[{Ramaswamy et~al.(2005)Ramaswamy, Cutright and
  Qammar}]{ramaswamy2005control}
\bibinfo{author}{Ramaswamy, S.}, \bibinfo{author}{Cutright, T.},
  \bibinfo{author}{Qammar, H.}, \bibinfo{year}{2005}.
\newblock \bibinfo{title}{Control of a continuous bioreactor using model
  predictive control}.
\newblock \bibinfo{journal}{Process Biochemistry} \bibinfo{volume}{40},
  \bibinfo{pages}{2763--2770}.
\bibitem[{Rao and Rawlings(2002)}]{rao2002constrained}
\bibinfo{author}{Rao, C.V.}, \bibinfo{author}{Rawlings, J.B.},
  \bibinfo{year}{2002}.
\newblock \bibinfo{title}{{Constrained process monitoring: Moving-horizon
  approach}}.
\newblock \bibinfo{journal}{AIChE journal} \bibinfo{volume}{48},
  \bibinfo{pages}{97--109}.
\bibitem[{Rawlings et~al.(2017)Rawlings, Mayne and Diehl}]{rawlings2017model}
\bibinfo{author}{Rawlings, J.B.}, \bibinfo{author}{Mayne, D.Q.},
  \bibinfo{author}{Diehl, M.}, \bibinfo{year}{2017}.
\newblock \bibinfo{title}{{Model predictive control: theory, computation, and
  design}}. volume~\bibinfo{volume}{2}.
\newblock \bibinfo{publisher}{Nob Hill Publishing Madison, WI}.
\bibitem[{del Rio-Chanona et~al.(2016)del Rio-Chanona, Zhang and
  Vassiliadis}]{del2016model}
\bibinfo{author}{del Rio-Chanona, E.A.}, \bibinfo{author}{Zhang, D.},
  \bibinfo{author}{Vassiliadis, V.S.}, \bibinfo{year}{2016}.
\newblock \bibinfo{title}{{Model-based real-time optimisation of a fed-batch
  cyanobacterial hydrogen production process using economic model predictive
  control strategy}}.
\newblock \bibinfo{journal}{Chemical engineering science}
  \bibinfo{volume}{142}, \bibinfo{pages}{289--298}.
\bibitem[{Rivadeneira et~al.(2017)Rivadeneira, Ferramosca and
  Gonz{\'a}lez}]{rivadeneira2017control}
\bibinfo{author}{Rivadeneira, P.S.}, \bibinfo{author}{Ferramosca, A.},
  \bibinfo{author}{Gonz{\'a}lez, A.H.}, \bibinfo{year}{2017}.
\newblock \bibinfo{title}{Control strategies for nonzero set-point regulation
  of linear impulsive systems}.
\newblock \bibinfo{journal}{IEEE Transactions on Automatic Control}
  \bibinfo{volume}{63}, \bibinfo{pages}{2994--3001}.
\bibitem[{Sawatzki et~al.(2018)Sawatzki, Hans, Narayanan, Haby, Krausch,
  Sokolov, Glauche, Riedel, Neubauer and
  Cruz~Bournazou}]{sawatzki2018accelerated}
\bibinfo{author}{Sawatzki, A.}, \bibinfo{author}{Hans, S.},
  \bibinfo{author}{Narayanan, H.}, \bibinfo{author}{Haby, B.},
  \bibinfo{author}{Krausch, N.}, \bibinfo{author}{Sokolov, M.},
  \bibinfo{author}{Glauche, F.}, \bibinfo{author}{Riedel, S.L.},
  \bibinfo{author}{Neubauer, P.}, \bibinfo{author}{Cruz~Bournazou, M.N.},
  \bibinfo{year}{2018}.
\newblock \bibinfo{title}{{Accelerated bioprocess development of
  endopolygalacturonase-production with saccharomyces cerevisiae using
  multivariate prediction in a 48 mini-bioreactor automated platform}}.
\newblock \bibinfo{journal}{Bioengineering} \bibinfo{volume}{5},
  \bibinfo{pages}{101}.
\bibitem[{Schmideder et~al.(2016)Schmideder, Cremer and
  Weuster-Botz}]{schmideder2016parallel}
\bibinfo{author}{Schmideder, A.}, \bibinfo{author}{Cremer, J.H.},
  \bibinfo{author}{Weuster-Botz, D.}, \bibinfo{year}{2016}.
\newblock \bibinfo{title}{{Parallel steady state studies on a milliliter scale
  accelerate fed-batch bioprocess design for recombinant protein production
  with Escherichia coli}}.
\newblock \bibinfo{journal}{Biotechnology progress} \bibinfo{volume}{32},
  \bibinfo{pages}{1426--1435}.
\bibitem[{Shen et~al.(2008)Shen, Wang, Feng and Xiu}]{shen2008bilevel}
\bibinfo{author}{Shen, L.}, \bibinfo{author}{Wang, Y.}, \bibinfo{author}{Feng,
  E.}, \bibinfo{author}{Xiu, Z.}, \bibinfo{year}{2008}.
\newblock \bibinfo{title}{Bilevel parameters identification for the multi-stage
  nonlinear impulsive system in microorganisms fed-batch cultures}.
\newblock \bibinfo{journal}{Nonlinear Analysis: Real World Applications}
  \bibinfo{volume}{9}, \bibinfo{pages}{1068--1077}.
\bibitem[{Sopasakis et~al.(2014)Sopasakis, Patrinos, Sarimveis and
  Bemporad}]{sopasakis2014model}
\bibinfo{author}{Sopasakis, P.}, \bibinfo{author}{Patrinos, P.},
  \bibinfo{author}{Sarimveis, H.}, \bibinfo{author}{Bemporad, A.},
  \bibinfo{year}{2014}.
\newblock \bibinfo{title}{Model predictive control for linear impulsive
  systems}.
\newblock \bibinfo{journal}{IEEE Transactions on Automatic Control}
  \bibinfo{volume}{60}, \bibinfo{pages}{2277--2282}.
\bibitem[{Tai et~al.(2015)Tai, Ly, Leung and Nayar}]{tai2015efficient}
\bibinfo{author}{Tai, M.}, \bibinfo{author}{Ly, A.}, \bibinfo{author}{Leung,
  I.}, \bibinfo{author}{Nayar, G.}, \bibinfo{year}{2015}.
\newblock \bibinfo{title}{{Efficient high-throughput biological process
  characterization: Definitive screening design with the Ambr250 bioreactor
  system}}.
\newblock \bibinfo{journal}{Biotechnology progress} \bibinfo{volume}{31},
  \bibinfo{pages}{1388--1395}.
\bibitem[{Tebbani et~al.(2008)Tebbani, Dumur and Hafidi}]{tebbani2008open}
\bibinfo{author}{Tebbani, S.}, \bibinfo{author}{Dumur, D.},
  \bibinfo{author}{Hafidi, G.}, \bibinfo{year}{2008}.
\newblock \bibinfo{title}{{Open-loop optimization and trajectory tracking of a
  fed-batch bioreactor}}.
\newblock \bibinfo{journal}{Chemical Engineering and Processing: Process
  Intensification} \bibinfo{volume}{47}, \bibinfo{pages}{1933--1941}.
\bibitem[{Thompson et~al.(2009)Thompson, McAuley and
  McLellan}]{thompson2009parameter}
\bibinfo{author}{Thompson, D.E.}, \bibinfo{author}{McAuley, K.B.},
  \bibinfo{author}{McLellan, P.J.}, \bibinfo{year}{2009}.
\newblock \bibinfo{title}{{Parameter estimation in a simplified MWD model for
  HDPE produced by a Ziegler-Natta catalyst}}.
\newblock \bibinfo{journal}{Macromolecular Reaction Engineering}
  \bibinfo{volume}{3}, \bibinfo{pages}{160--177}.
\bibitem[{Villa-Tamayo and Rivadeneira(2020)}]{villa2020adaptive}
\bibinfo{author}{Villa-Tamayo, M.F.}, \bibinfo{author}{Rivadeneira, P.S.},
  \bibinfo{year}{2020}.
\newblock \bibinfo{title}{Adaptive impulsive offset-free mpc to handle
  parameter variations for type 1 diabetes treatment}.
\newblock \bibinfo{journal}{Industrial \& Engineering Chemistry Research}
  \bibinfo{volume}{59}, \bibinfo{pages}{5865--5876}.
\bibitem[{W{\"a}chter and Biegler(2006)}]{wachter2006implementation}
\bibinfo{author}{W{\"a}chter, A.}, \bibinfo{author}{Biegler, L.T.},
  \bibinfo{year}{2006}.
\newblock \bibinfo{title}{{On the implementation of an interior-point filter
  line-search algorithm for large-scale nonlinear programming}}.
\newblock \bibinfo{journal}{Mathematical programming} \bibinfo{volume}{106},
  \bibinfo{pages}{25--57}.
\bibitem[{Yang(2001)}]{yang2001impulsive}
\bibinfo{author}{Yang, T.}, \bibinfo{year}{2001}.
\newblock \bibinfo{title}{Impulsive control theory}. volume
  \bibinfo{volume}{272}.
\newblock \bibinfo{publisher}{Springer Science \& Business Media}.
\bibitem[{Yang et~al.(2019)Yang, Peng, Lv and Li}]{yang2019recent}
\bibinfo{author}{Yang, X.}, \bibinfo{author}{Peng, D.}, \bibinfo{author}{Lv,
  X.}, \bibinfo{author}{Li, X.}, \bibinfo{year}{2019}.
\newblock \bibinfo{title}{Recent progress in impulsive control systems}.
\newblock \bibinfo{journal}{Mathematics and Computers in Simulation}
  \bibinfo{volume}{155}, \bibinfo{pages}{244--268}.
\bibitem[{Yin and Liu(2017)}]{yin2017distributed}
\bibinfo{author}{Yin, X.}, \bibinfo{author}{Liu, J.}, \bibinfo{year}{2017}.
\newblock \bibinfo{title}{Distributed moving horizon state estimation of
  two-time-scale nonlinear systems}.
\newblock \bibinfo{journal}{Automatica} \bibinfo{volume}{79},
  \bibinfo{pages}{152--161}.

\end{thebibliography}





\end{document}